\documentclass[letterpaper,11pt]{amsart}
\usepackage{times}
\usepackage[margin=1.20in]{geometry}
\usepackage{graphicx}
\usepackage[hyperfootnotes=false]{hyperref}
\usepackage{times}
\usepackage[margin=1.20in]{geometry}
\usepackage{graphicx}
\usepackage[dvipsnames]{xcolor}
\let\Hor\H

\usepackage{amssymb}
\theoremstyle{plain}
\newtheorem{theorem}{Theorem}%[section]

\theoremstyle{definition}

\theoremstyle{remark}

\newcommand{\suppress}[1]{}

\numberwithin{equation}{section}

\newcommand{\be}{\begin{equation}}
\newcommand{\ee}{\end{equation}}
\newcommand{\bea}{\begin{eqnarray}}
\newcommand{\eea}{\end{eqnarray}}
\newcommand{\bean}{\begin{eqnarray*}}
\newcommand{\eean}{\end{eqnarray*}}

\usepackage{color}

\title [The quantum computer puzzle] {{\LARGE The quantum computer puzzle}\\ \medskip {\Large (Expanded Version)}}

\author[Gil Kalai]{\Large Gil Kalai}\address{Hebrew University of Jerusalem and Yale
University} \email{kalai@math.huji.ac.il}\thanks{Work
supported in part by ERC advanced grant 320924, BSF grant 2006066, and NSF grant DMS-1300120.
This is an expanded version of my paper 
in the 
\href {http://www.ams.org/journals/notices/201605/}
{{\it Notices of the AMS}, May 2016.}
Sections 1--6 are slightly expanded and references and few comments are added. Sections 6.2, 7, and 8 are new.
}

\begin{document}

\clearpage

\maketitle

\begin {quotation}
{\color {purple} %\large
%{\large
%...chaque nouvelle possibilit\'e qu'a l'existence,
%m\^eme celle qui est la moins probable, transforme l'existence
%tout enti\'ere,
Any new
possibility that existence acquires, even the least
likely, transforms everything about existence.
}
\end {quotation}
\begin {flushright}
{\color {purple}  
%{\large 
Milan Kundera -- {\em Slowness}}

%\footnote{ 
%{\color {red}
%}
%-- from {\it Slowness} by Milan Kundera (translated by Linda Asher.)%}
%}

\end {flushright}

%\hskip{0.7cm}

\indent
Quantum computers
are hypothetical devices, based on quantum physics,
that would enable us to perform certain computations
hundreds of orders of magnitude  faster than digital computers \cite {Fey82,Deu85}.
This feature is
coined as ``quantum supremacy'' \cite {Pre12}
and one aspect or another of such quantum computational supremacy might be brought about in
experiments in the near future:
by implementing quantum error-correction,  systems of non-interacting bosons,
exotic new phases of matter called anyons,  quantum annealing, or in various
other ways.
We concentrate in this paper on
the model of a universal quantum computer,
which allows the full computational potential of quantum systems,
and on the  restricted model, called ``BosonSampling,''
based on non-interacting bosons.

A main concern regarding the feasibility
of quantum computers is that quantum systems are
inherently noisy: we cannot accurately control them, and we
cannot accurately describe them.
We will describe an optimistic hypothesis
of quantum noise that would allow
quantum computing and a pessimistic hypothesis that wouldn't.
The {\it quantum computer puzzle} is  deciding between these two
hypotheses.\footnote{ More broadly, the quantum computer puzzle is the question of whether
quantum computers are possible and whether quantum supremacy is a real phenomenon.
It is common to regard quantum noise as the crux of the matter
but there are other views.}
We list some remarkable consequences of the optimistic hypothesis, giving strong
reasons for the intensive efforts to build
quantum computers, as well as good reasons for suspecting that this might not be possible.
For systems of non-interacting bosons,
we explain how quantum supremacy achieved without noise
is replaced, in the presence of noise, by a very low yet fascinating
computational power (based on
%\footnote {Based on
%G. Kalai and G. Kindler, Gaussian noise sensitivity and BosonSampling, arXiv:1409.3093.}
Kalai and Kindler \cite {KalKin14}).
Finally, based on the pessimistic hypothesis,
we make seventeen predictions about
quantum physics and computation (based on, Kalai \cite {Kal11},
and a subsequent Internet debate with Aram Harrow and others\footnote {The 2012 debate \cite {KalHar12} took
place %in 2012 
over Lipton and Regan's blog ``G\"odel's lost letter
and NP $\ne$ P.'' There were more than a thousand comments by
over a hundred participants, renowned scientists along with non-experts. 
Some further discussions with Peter Shor and others
took place on the author's blog ``Combinatorics and more'' and in other places.}).

Are quantum computers feasible? Is quantum supremacy possible? My expectation is that the pessimistic
hypothesis will prevail, leading to a negative answer. Rather than regard
this possibility as an unfortunate failure
that impedes the progress of humanity, I believe that the failure of
quantum supremacy itself leads
to important consequences for quantum physics, the theory of
computing, and mathematics. Some of these will be explored here.

\subsection *{The essence of my point of view}

Here is a brief summary of the author's pessimistic
point of view as explained in the paper:
Understanding
quantum computers in the presence of noise requires
consideration of behavior at different scales. In the small
scale, standard models of noise from the mid-90s are
suitable, and quantum evolutions and states described
by them manifest a very low-level computational power.
This small-scale behavior has far-reaching consequences
for the behavior of noisy quantum systems at larger
scales. On the one hand, it does not allow reaching the
starting points for quantum fault tolerance and quantum
supremacy, making them both impossible at all scales.
On the other hand, it leads to novel implicit ways for
modeling noise at larger scales and to various predictions
on the behavior of noisy quantum systems.
%Understanding quantum computers in the presence of noise should be carried out in two scales.
%%%Let me start by summarizing  the picture drawn in Sections 5,6, and 7.
%In the small scale, modeling noise according to the standard
%models of noise as done since the mid 90s (and earlier)
%is perfectly suitable.
%%%Standard noise models for evolutions and states of quantum devices in the small scale
%%%can be
%%%modelled by the standard noise models and this
%Quantum evolutions and quantum states
%as described in the small scale by the standard noise models
%%%%can be described by low degree polynomials and therefore these systems
%manifest a very low-level computational power (Section 5), which does not
%%%lead to  a description by low degree polynomials
%%%which manifests very low computational power that do
%allow reaching the starting points for quantum fault-tolerance and quantum supremacy.
%%%% to emerge when we move to somewhat larger scales.
%%%%The inability to reach the starting point for quantum fault-tolerance in the small scale
%This small-scale behavior has far-reaching consequences for the behavior of noisy quantum systems
%in the medium and large scale. In particular, it leads to novel implicit
%ways for modeling noise (Sections 6),
%and to various predictions on the behavior of noisy quantum systems (Sections 5, 6 and 7).
%. which express the additional property of
%``no quantum fault-tolerance.''
The small-scale behavior of noisy quantum systems
does  allow, for larger scales, the creation of robust {\it classical}
information and computation.

This point of view is expected to be tested in various
experimental efforts to demonstrate quantum supremacy in the next few
years.\footnote{This is a good place
to emphasize that the
purpose of my work is not to give a
mathematical proof that quantum computers are not possible.
Rather, the two-scale mathematical modeling of noisy quantum systems could be part
of a {\it scientific proof} that quantum supremacy is not a real phenomenon and that
quantum computers are not possible.}

%As convincing experimental evidence for quantum supremacy are expected by many
%\subsubsection *{The gap in intuitions regarding scaling}
%The pessimistic and optimistic
%hypotheses reflect different intuitions
%for the %engineering
%difficulties of scaling up systems.
%The optimistic hypothesis relies on the belief that
%scaling up an engineering device based on $n$ elements
%represents polylogarithmic or polynomial-scale difficulty
%rather than exponential difficulty.
%In view of the picture drawn in Section 5,
%the pessimistic hypothesis is based on the following alternative:
%Scaling up an engineering device based on $n$ elements which perform asymptotically a task in
%a very low-level complexity class will fail well before the device
%demonstrates full classic computational powers or superior computational powers.

\section {The vision of quantum computers and quantum supremacy}

\label {int}

\subsection {Circuits and quantum circuits}

The basic memory component in classical computing is a ``bit,'' which
can be in two states, ``0'' or ``1.'' A computer (or a circuit) has
$n$ bits and it can perform certain logical operations  on them. The NOT gate, acting on a single bit, and the AND
gate, acting on two bits, suffice for {\it universal} classical computing.
This means that a computation based on another collection of logical gates,
each acting on a bounded number of bits,
can be replaced by a computation based only on NOT and AND.
Classical circuits equipped with random bits lead to {\it randomized algorithms}, which are both practically useful
and theoretically important.

Quantum computers allow the creation of probability
distributions that are well beyond the reach of classical computers with access to random bits.
A qubit is a piece of quantum memory.
The state of a qubit can be described by a unit
vector in a two-dimensional complex Hilbert space $H$.
For example, a basis for $H$ %to this Hilbert space
can correspond to two energy levels of the hydrogen
atom, or to horizontal and vertical polarizations of a photon.
Quantum mechanics allows the qubit to be in a {\it superposition} of the basis vectors, described by
an arbitrary unit vector in $H$.
The memory of a
quantum computer (``quantum circuit'') consists of $n$ qubits. Let $H_k$ be the two-dimensional Hilbert space
associated with the
$k$th qubit.
The state of the entire memory of $n$ qubits is described by
a unit vector in the tensor product $H_1 \otimes H_2 \otimes \cdots \otimes H_n$.
We can put one
or two
qubits through {\it gates} representing
unitary transformations acting on
the corresponding two- or four-dimensional Hilbert spaces, and as for classical computers, there is a
small list of gates sufficient for  universal quantum computing.\footnote{
The notion of universality in the quantum case relies on
the Solovay--Kitaev theorem \cite {NC00}[App. 3].   % which is a crucial result
%also to other aspects of quantum computing.
Several universal classes of quantum gates are described in \cite {NC00}[Ch. 4.5].}
%Each step in the computation process consists of applying a unitary transformation
%on the large $2^n$-dimensional Hilbert space, namely,
%applying a gate on one or two qubits, tensored with the identity transformation
%on all other qubits.
At the end of the computation process, the state of the entire computer can be {\it measured}, giving a probability
distribution on 0--1 vectors of length $n$.

To review, classical and quantum computation processes are similar.
In the classical case we have a computer memory of $n$ bits, and
every computation step consists of applying logical Boolean gates to
one or two bits (without changing the values of other bits).
We can also add randomization by allowing the value of
some bits to be set to 0 with probability 1/2 and to 1 with probability 1/2.
For the quantum case, the state of the $n$-qubit computer is a unit vector
in a large $2^n$-dimensional complex vector space.
Each computation step is a unitary transformation (chosen from a list of quantum gates)
on the Hilbert space describing the qubits on which the gate acts,
tensored with the identity transformation
on all other qubits.

A few words on the connection between the mathematical model of quantum circuits and quantum physics:
in quantum physics, states and
their evolutions (the way they change in time)
are governed by the Schr\"odinger equation.
A solution of the
Schr\"odinger equation can be described
as a unitary process on a Hilbert space and
quantum computing processes of the kind we just described form a large class of such quantum evolutions.

\subsection {A very brief tour of computational complexity}

\begin{figure}
\centering
\includegraphics[scale=0.6]{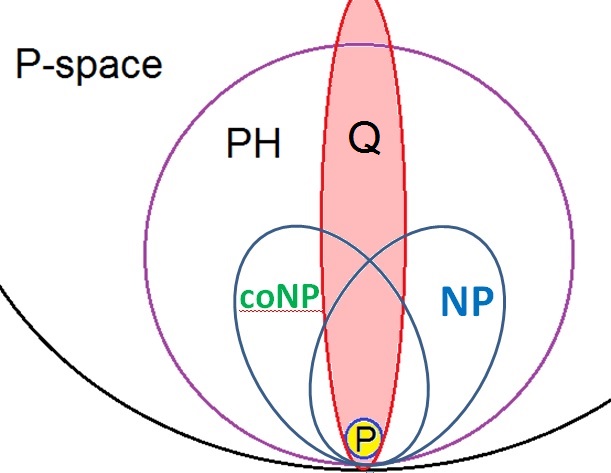}
\caption{The (conjectured) view of some main computational complexity classes. The red ellipse represents
efficient quantum algorithms.}
\label{fig:0}
\end{figure}

Computational complexity
is the theory of {\it efficient computations}, where ``efficient''
is an asymptotic notion referring to situations where
the number of computation steps (``time'') is at most a polynomial in the number of input bits.
For modern books on computational complexity, the reader is referred to
Goldreich \cite {Gol10,Gol08} and Arora and Barak \cite {AB09}.

The complexity class {\bf P} is the class of algorithms that can be performed using a polynomial
number of steps in the size of the input.
The complexity class {\bf NP} refers to non-deterministic polynomial time.
Roughly speaking, it refers to questions
where we can {\em provably} perform the task in a polynomial number of
operations in the input size, provided we are given a
certain polynomial-size ``hint'' of the solution.
An algorithmic task $A$ is {\bf NP}-hard if a subroutine for solving
$A$ allows solving any problem in {\bf NP}
in a polynomial number of steps. An {\bf NP}-complete
problem is an {\bf NP}-hard problem in {\bf NP}.
Examples of {\bf NP}-complete problems are: to decide if a graph
has a Hamiltonian cycle, or to decide if a Boolean formula
has a satisfying assignment. A  problem is in {\bf coNP} if its
complement is in {\bf NP}. For example, to decide if a graph does
not have a Hamiltonian cycle is in {\bf coNP}.

A useful analog is to think about the gap between {\bf NP} and {\bf P}
as similar to the gap between finding a proof of a theorem and verifying
that a given proof of the theorem is correct.
{\bf P}, {\bf NP}, and {\bf coNP} are three of the lowest
computational complexity classes in
the {\em polynomial hierarchy {\bf PH}}, which is a countable sequence of such
classes, and there is a rich theory of complexity classes beyond {\bf PH}.

There are intermediate problems between {\bf P} and {\bf NP}. Factoring an $n$-digit integer
is not known to be in {\bf P}, as the best algorithms
are exponential in the cube root of the number of digits. Factoring is in {\bf NP}, but it is unlikely
that factoring is {\bf NP}-complete.
Shor's famous algorithm shows that quantum computers can factor $n$-digit integers efficiently --
in $\sim n^2$ steps! Quantum computers are not known to be able to solve
{\bf NP}-complete problems efficiently, and there are good reasons to think that they cannot. However,
quantum computers can
efficiently perform certain computational tasks
beyond {\bf NP}. The class of decision
problems (algorithmic tasks with a yes/no answer)
that quantum computers can efficiently solve is denoted by {\bf BQP}.

Two comments: first, our understanding of the world of computational complexity depends on a
whole array of conjectures: {\bf NP} $\ne$ {\bf P} is the most famous one, and a stronger conjecture
asserts that {\bf PH} does not collapse,
namely, that there is a strict inclusion between the computational
complexity classes defining the polynomial hierarchy.
Second, computational complexity insights,
while asymptotic, strongly apply to finite and small algorithmic tasks. The following example will be 
important for our analysis: recall that the Ramsey number $R(n,n)$ is the smallest $m$ such that
for every coloring of the edges of a complete graph on $m$ vertices with two colors,
there is a complete graph on $n$ vertices all of whose edges are colored with the same color. 
Paul Erd\Hor{o}s famously claimed that finding the
value of the Ramsey function $R(n,n)$ for $n=6$ is well beyond mankind's ability.
This statement is supported by  computational complexity insights that consider the
difficulty of computations as $n \to \infty$,
while not directly implied by them.

\section {Noise}
\label {s:noise}
\subsection {Noise and fault-tolerant computation}

The main concern
regarding the feasibility of quantum computers has always been that
quantum systems are
inherently noisy: we cannot accurately control them, and we
cannot accurately describe them. The concern regarding noise in quantum systems as a major
obstacle to quantum computers was put forward
in the mid-90s by Landauer \cite{lan95}, Unruh \cite{Unr95}, and others.\footnote{A few additional 
papers (among many) expressing  concerns regarding the feasibility 
of quantum computers or studying critically such concerns 
are \cite{HaRa96,Pre98,Lev03,Gol04,Ali04,Aar04,Kal05,Dya07,Ali13,Pre13}.} 
To overcome this difficulty, a
theory of quantum fault-tolerant computation based on quantum error-correcting codes
was developed \cite{Sho95,Ste96,Sho96,AhaBen97,Kit97,Got97,KLZ98}; see 
also \cite {NC00}[Ch. 10].\footnote{The study of quantum error-correcting codes is a fascinating addition to 
the classical theory of error-correcting codes which account for some of the 
most important practical applications of mathematics.}
Fault-tolerant computation refers to computation in the presence of errors. The
basic idea is to represent (or ``encode'') a single piece of information (a bit in the classical case
or a qubit in the quantum case) by a large number of physical components so as to ensure
that the computation is robust even if some of these physical components are faulty.

What is noise? Solutions of the Schr\"odinger equation (``quantum evolutions'') can be regarded as unitary
processes on Hilbert spaces.
Mathematically speaking, the study of noisy quantum systems is the
study of {\it pairs} of Hilbert spaces $(H, H')$, $H \subset H'$,
and a unitary process on the larger Hilbert space $H'$.
Noise refers to the general effect of neglecting degrees
of freedom, namely, approximating the process on a large Hilbert space
by a process on the small Hilbert space. For controlled quantum systems and,
in particular, quantum computers, $H$ represents the controlled part of the system,  and
the large unitary process on $H'$  represents, in addition to an ``intended''
controlled evolution on $H$,
also the uncontrolled effects of the {\em environment}.
The study of noise is relevant, not only to controlled
quantum systems, but also to many other aspects of quantum physics.

A second, mathematically equivalent way
to view noisy states and noisy evolutions is to stay with the original Hilbert space $H$, but to
consider a mathematically larger class of states and operations.
In this view, the state of a noisy qubit is described as a classical probability distribution on
unit vectors of the associated Hilbert spaces. Such states are referred to as {\it mixed states}.
It is convenient to think about the following form of noise, called {\it depolarizing noise}:
in every computer cycle a qubit is not affected with probability $1-p$, and, with probability
$p$, it turns into the \emph{maximal entropy mixed state}, i.e.,
the average of all unit vectors in the associated Hilbert
space. In this example, $p$ is the error rate, and, more generally,
the error rate can be defined as the probability that a qubit is corrupted
at a computation step conditioned on it having survived up to this step.

\subsection {Two alternatives for noisy quantum systems}

\begin{figure}
\centering
\includegraphics[scale=0.3]{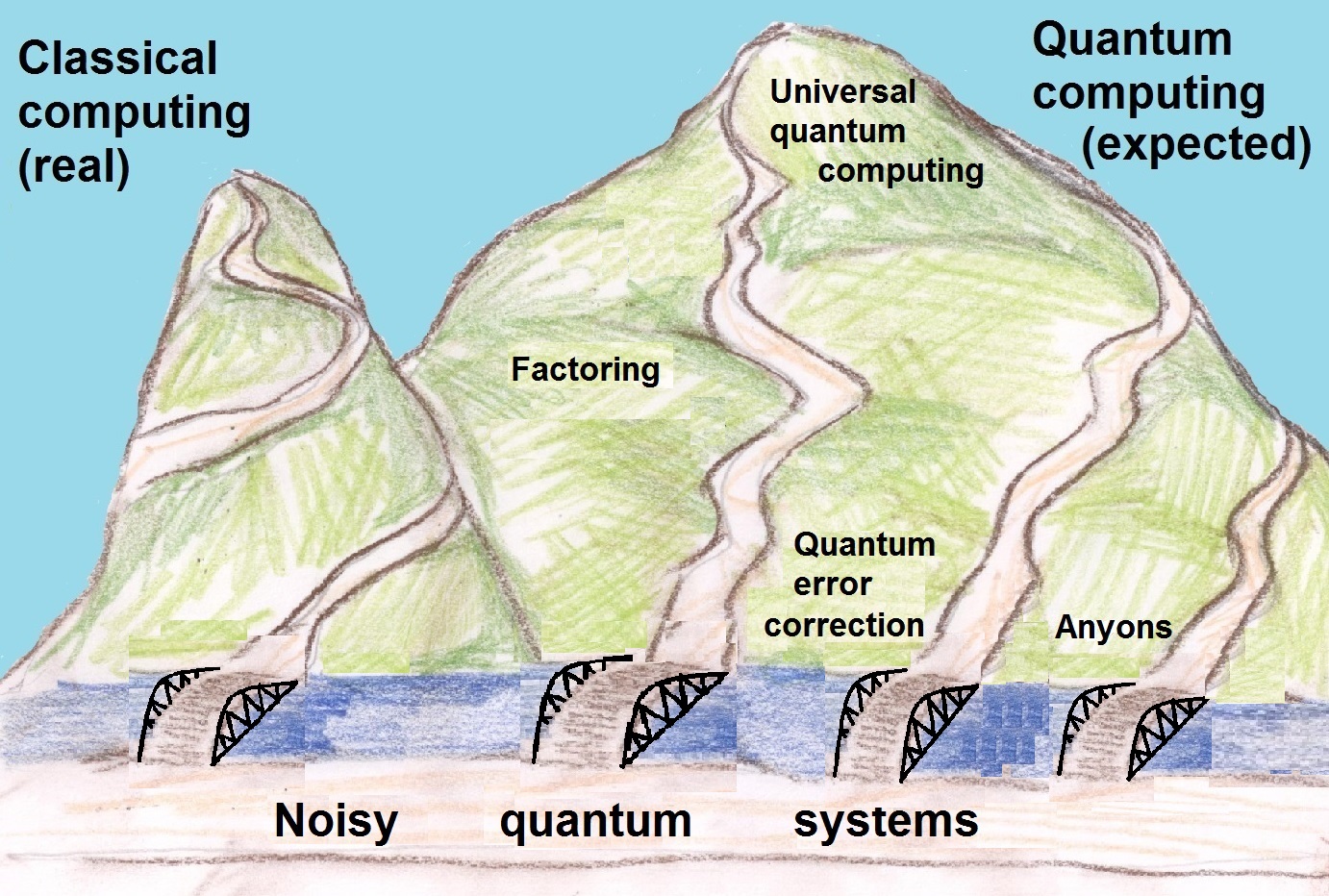}
\caption{{\bf The optimistic hypothesis:}  Classical fault-tolerance mechanisms
can be extended, via quantum error-correction, allowing robust quantum information and
computationally superior quantum computation.
 Drawing by Neta Kalai.}
\label{fig:opt}
\end{figure}

The quantum computer puzzle is, in a nutshell, deciding between two hypotheses
regarding properties of noisy quantum circuits,
the {\it optimistic hypothesis} and the {\it pessimistic hypothesis}.

\medskip

{\bf Optimistic hypothesis:} It is possible to realize universal quantum
circuits with a small bounded error level regardless of the number of qubits.
The effort required to obtain a bounded error level for universal quantum circuits
increases
moderately with the number of qubits.
Therefore, large-scale fault-tolerant quantum computers are possible.

\medskip

{\bf Pessimistic hypothesis:}  The error rate
in every realization of  universal quantum circuits scales up (at least) linearly with the number of qubits.
The effort required to obtain a bounded error level
for any implementation of universal quantum circuits
increases (at least) exponentially with the number of qubits. Thus, quantum computers are not possible.

\medskip

Some explanations: for the optimistic hypothesis, we note that
the main theorem of quantum fault-tolerance asserts that (under some natural conditions on the noise),
if we can realize universal quantum circuits
with a sufficiently small error rate (where the threshold is roughly between 0.001 and 0.01,) then
quantum fault-tolerance and hence universal quantum computing are possible.
For the pessimistic hypothesis, when we say
that the rate of noise per qubit scales up linearly with the number of qubits
we mean that when we double the number of qubits in the circuit the probability for a single qubit
to be corrupted in a small time interval doubles.
The pessimistic hypothesis
does not require new modeling of the noise of universal quantum circuits, and it is just based on
a different assumption on the rate of noise.
However, for more general noisy quantum systems, it leads to interesting predictions and modeling, and may  lead to useful
computational tools.
We emphasize that both hypotheses
are assertions about physics (or physical reality),
not about mathematics, and both of the hypotheses represent scenarios
that are compatible with quantum mechanics.%\footnote {Some researchers regard noise as
%solely an engineering issue, and others even posit that noise
%does not have any objective meaning.
%These views raises interesting
%conceptual issues and are related to the question if
%quantum supremacy is a real phenomenon.}

The constants are important and the pessimistic view regarding
quantum supremacy holds that every realization
of universal quantum circuits
will fail for a handful of qubits,
long before any quantum supremacy effect is witnessed, and long before quantum fault-tolerance is possible.
The failure to reach universal quantum circuits for a small
number of qubits, and to manifest quantum supremacy
for small quantum systems, is
crucial for the pessimistic hypothesis, and
Erd\Hor{o}s's statement about $R(6,6)$ is a good analogy for this expected behavior.

Both on the technical and conceptual  levels we see here what we call a ``wide-gap
dichotomy.'' On the technical level, we have a gap between small constant error rate per qubit for the
optimistic view, and a linear increase of rate per qubit (in terms of the number
of qubits in the circuit) on the pessimistic side. We
also have a gap between the ability to achieve large-scale
quantum computers on the optimistic side, and the failure of universal
quantum circuits already for a handful of qubits
on the pessimistic side.
On the conceptual level, the optimistic hypothesis asserts that quantum mechanics
allows superior computational powers,
while the pessimistic hypothesis asserts that quantum systems without  specific mechanisms for
robust classical information that leads only to classical
computing are actually computationally inferior.
We will come back to both aspects of this wide-gap dichotomy.

\subsection {Potential experimental support for quantum supremacy}

A definite demonstration of quantum supremacy of controlled
quantum systems, namely, building quantum systems
that outperform, even for specific computational tasks, classical computers,
or a definite demonstration of quantum error-correction, will falsify the pessimistic hypothesis
and will lend strong support to the optimistic hypothesis. (The optimistic hypothesis will be
completely verified with full-fledged universal quantum computers.)
There are several ways people are planning, in the next few years,
to demonstrate quantum
supremacy or the feasibility of quantum fault-tolerance.\footnote{Some researchers
refer to an empirical demonstration
of quantum supremacy as ``imminent.''}

%e.g., via a small programmable
%quantum computers based on 40 or more high quality qubits.}

\begin {enumerate}
\item
Attempts to create small universal quantum circuits with up to ``a few tens of qubits.''
\item
Attempts to create stable logical qubits based on surface codes.
\item
Attempts to have BosonSampling for 10--50 bosons.
\item
Attempts to create stable qubits based on anyonic states.
\item
Attempts to demonstrate quantum speed-up based on quantum annealing.
\end {enumerate}

Each of attempts (1)--(4) represents
many different experimental directions carried out mainly in academic
institutions (and research centers of large companies
like IBM, Microsoft, and Google),\footnote{A new company QCI, whose long-term goal is to develop quantum computers
based on the model of quantum circuits and quantum error-correction was recently
established by a group of researchers from Yale.} while (5)
represents an attempt by a commercial company: D-wave.\footnote{D-wave is attempting to
demonstrate quantum speedup for {\bf NP}-hard optimization problems, and even to compute Ramsey numbers.}
There are many different avenues for realizing qubits, of which
ion-trapped qubits and superconducting qubits are perhaps the leading ones, and there are several
groups attempting to demonstrate stable logical qubits via quantum error-correction. %in the next few years.
Quantum supremacy via nonabelian anyons stands out
as a very different direction based on exotic new phases of matter and very deep mathematical and physical issues.
BosonSampling (see Section \ref {s:bs})
stands out in the quest to demonstrate quantum supremacy for narrow physical systems
without offering further practical fruits.

The pessimistic hypothesis
predicts a decisive failure for {\it all} of these attempts to demonstrate quantum supremacy,
or very stable logical qubits, and also that
this failure will be witnessed for small systems.
A reader may ask how the optimistic hypothesis can be falsified, beyond repeated
failures to demonstrate universal quantum computers or partial steps toward them such as those listed above.
My view is that the optimistic hypothesis could be largely falsified
if we can understand the absence of quantum supremacy and quantum error-correction as a physical principle
with prediction power that goes beyond these repeated failures -- both in providing more detailed predictions about
these failures themselves (such as scaling up of errors, correlations between errors, etc.) and
in providing predictions about other natural quantum systems.
Mathematical modeling
of noisy quantum systems based on the pessimistic hypothesis is valuable, not only if it represents
a general  physical principle, but also if it represents temporary
technological difficulties or if it applies to limited classes of quantum systems.

\section {BosonSampling}
\label{s:bs}

Quantum computers would allow the creation of probability
distributions that are beyond the reach of classical computers with access to random bits.
This is manifested by ``BosonSampling,''
a  class of probability distributions representing a collection of non-interacting bosons,
that quantum computers can efficiently create. It is a
restricted subset of distributions compared to the class of distributions that a
universal quantum computer can produce, and it is not known
if BosonSampling distributions can be used for efficient integer factoring or
 other ``useful'' algorithms.
BosonSampling was introduced by Troyansky and Tishby in 1996 and was intensively studied
by Aaronson and Arkhipov \cite {AaAr13}, who offered it as a
quick path for experimentally showing that quantum supremacy is a real phenomenon.

Given an $n$ by $n $ matrix $A$, let $det (A)$ denote the determinant of $A$, and $per (A)$
denote the permanent of $A$. Thus $det (A) = \sum_{ \pi \in S_n} sgn(\pi) \prod_{i=1}^n a_{i \pi(i)}$,
and $per (A) = \sum_{ \pi \in S_n}  \prod_{i=1}^n a_{i \pi(i)}$.
Let $M$ be a complex $n \times m$ matrix, $m \ge n$.
Consider all ${m} \choose {n}$ subsets $S$ of $n$ columns, and for every subset
consider the corresponding $n \times n$ submatrix $A$.
The algorithmic task of sampling subsets $S$ of columns according to $|det (M')|^2$
is called FermionSampling.
Next consider all ${m+n-1} \choose {n}$ sub-multisets $S$ of $n$ columns (namely, allow columns to repeat),
and for every sub-multiset $S$ consider the corresponding $n \times n$ submatrix $A$ (with column $i$ repeating $r_i$ times).
BosonSampling
is the algorithmic task of sampling those multisets $S$ according to
$|per (A)|^2/(r_1!r_2!\cdots r_n!)$. Note that
the algorithmic task for BosonSampling and FermionSampling
is to sample according to a specified probability distribution.
This is not a decision problem, where the algorithmic task is to
provide a yes/no answer.

Let us demonstrate these notions by an example for $n=2$ and $m=3$.
The input is a $2 \times 3$ matrix:

\[ \left( \begin{array}{ccc}
1/\sqrt 3 & i/\sqrt 3 & 1/\sqrt 3 \\
0 & 1/\sqrt 2 & i/\sqrt 2 \end{array} \right)\]

The output for FermionSampling is a probability distribution on subsets of two columns, with probabilities
given according to absolute values of the square of determinants.
Here we have probability $1/6$ to columns $\{1,2\}$, probability $1/6$ to columns $\{1,3\}$, and
probability $4/6$ to columns $\{2,3\}$.
The output for BosonSampling is a probability distribution according to absolute values of the square of
permanents of sub-multisets of two columns.
Here, the probabilities are: $\{1,1\} \to 0$; $\{1,2\} \to 1/6$; $\{1,3\} \to 1/6$;
$\{2,2\} \to 2/6$; $\{2,3\} \to 0$; $\{3,3\} \to 2/6$.

FermionSampling describes the state of $n$ non-interacting fermions, where each individual
fermion is described as a superposition of $m$ ``modes.''  BosonSampling describes the state of $n$ non-interacting fermions,
where each individual fermion is described by $m$ modes. A few words about the physics:
fermions and bosons are the main building blocks of nature.
Fermions, such as electrons, quarks, protons, and neutrons, are particles characterized by Fermi--Dirac statistics.
Bosons, such as photons, gluons, and the Higgs boson, are particles characterized
 by Bose--Einstein statistics.

Moving to computational complexity, we note that
Gaussian elimination gives an efficient algorithm for computing determinants, but
computing permanents is very hard: it represents a computational complexity class,
called {\bf \#P} (in words, ``number {\bf P}'' or ``sharp {\bf P}''), that extends beyond
the entire polynomial hierarchy.
It is commonly believed that even quantum computers cannot efficiently compute permanents.
However, a quantum computer can efficiently create a bosonic (and a fermionic) state based on a matrix $M$,
and therefore perform efficiently both BosonSampling and FermionSampling.
A classical computer with access to random
bits can sample FermionSampling efficiently, but, as proved by
Aaronson and Arkhipov,
a classical computer with access to random bits
cannot perform BosonSampling
unless the polynomial hierarchy collapses! (See \cite{AaAr13,BJS11,TeDi04}.)

\section {Predictions from the optimistic hypothesis}
\label {s:r2d}

\subsubsection* {Barriers crossed}
Quantum computers would dramatically change our reality.

\begin {enumerate}

\item
A universal machine for creating quantum states and evolutions will be built.

\item
Complicated evolutions and states with global interactions,
markedly different from anything witnessed so far, will be created.

\item
It will be possible to experimentally time-reverse every quantum evolution.

\item
The noise will not respect symmetries of the state.

\item
There will be fantastic computational complexity consequences.

\item

Quantum computers will efficiently break most current public-key cryptosystems.
\end {enumerate}

Items 1--4 represent a vastly different experimental reality than that of today, and items 5 and 6 represent
a vastly different computational reality.\footnote{Recently, in response to the last item,
the NSA (U.S.- National Security Agency)
publicly set a goal of ``a transition to quantum
resistant algorithms in the not too distant future.''}

\subsubsection*  {Magnitude of improvements}
It is often claimed
that quantum computers
could perform in a few hours certain computations
that take longer than the lifetime of the universe on a classical computer!
Indeed, it is useful to examine not only things that were previously impossible and which are now made
possible by a new technology,
but also the improvement in terms of orders of magnitude for tasks that could have been achieved
by the old technology.
Quantum computers represent enormous, unprecedented, order-of-magnitude
improvement of controlled physical phenomena as well as of algorithms.
Nuclear weapons represent an improvement of 6--7 orders of magnitude over conventional ordinance:
the first atomic bomb was a million
times stronger than the most powerful (single) conventional bomb at the time.  The telegraph could deliver
a transatlantic message in a few seconds compared to
the previous three-month period.
This represents an (immense) improvement of 4--5 orders of magnitude.
Memory and speed of computers were improved by 10--12 orders of magnitude over several decades.
Breakthrough algorithms at the time of their discovery also
represented practical improvements
of no more than a few orders of magnitude.
Yet implementing BosonSampling with a hundred bosons represents
more than a hundred orders of magnitude of improvement compared to digital computers, and
a similar story can be told about a large-scale quantum computer applying Shor's
algorithm.\footnote{We note that quantum computers will not increase computational power
across the board (an increase of the kind witnessed by ``Moore's law'')
and that their applications are restricted and subtle.}

\subsubsection*
{Computations in quantum field theory}
Quantum electrodynamics (QED) computations allow one to describe various physical quantities
in terms of a power series
$$\sum c_k \alpha^k,$$
where $c_k$  is the contribution of Feynman's diagrams with $k$ loops, and $\alpha$
 is the fine structure constant (around 1/137). Quantum computers
will (likely\footnote {This plausible conjecture, which motivated quantum computers to start with,
is supported by the recent work of  Jordan, Lee, and  Preskill \cite {JLP14}, and is often taken
for granted. We note, however, that a rigorous mathematical framework for QED computations
is not yet available, and an efficient quantum algorithm
for these computations may require such a framework, or may serve as a major
step toward it.}) allow one to compute these terms and sums for large values of $k$
with  hundreds of digits of accuracy, similar to computations of
the digits of $e$ and $\pi$ on today's computers, even in regimes  where they have
no physical meaning!

\subsubsection* {My interpretation}

I regard the incredible consequences from the optimistic hypothesis
as solid indications
that quantum supremacy might be ``too good to be true,'' and
that the pessimistic hypothesis would prevail. Quantum computers would  change
reality
in unprecedented ways, both qualitatively and quantitatively,
and it is easier to believe that we will witness
substantial theoretical changes in modeling quantum noise than that we will witness
such dramatic changes in reality itself.

\section {BosonSampling meets reality}
\label {s:bs2}
\subsection {How does noisy BosonSampling behave?}

BosonSampling and noisy BosonSampling (i.e., BosonSampling in the presence of noise) exhibit
radically different behavior.
BosonSampling is based on $n$ non-interacting indistinguishable
bosons with $m$ modes. For noisy Boson Samplers
these bosons will not be perfectly non-interacting (accounting for one form of noise)
and will not be perfectly indistinguishable (accounting for another form of noise).
The same is true if we replace bosons by fermions everywhere.
The state of $n$ bosons with $m$ modes is represented by an algebraic variety
of decomposable symmetric tensors of real dimension $2mn$
in a huge relevant Hilbert space of dimension $2m^n$. For the fermion case this manifold is
simply the Grassmanian. The study of noisy BosonSampling in \cite {KalKin14} is based on a
general framework for the study of noise and
sensitivity to noise via Fourier expansion that
was introduced by Benjamini, Kalai, and Schramm \cite {BKS99};
see also \cite {GaSt14}.

We have already
discussed the rich theory of computational complexity classes
beyond {\bf P}, and there is also a rich theory below {\bf P}. One very low-level complexity class
consists of
computational tasks that can be carried out by bounded-depth
polynomial-size circuits.\footnote {For decision problems this class is referred to
as {\bf AC}$^{\bf 0}$.}
In this model the number of gates is, as before, at most
polynomial in the input side, but an additional severe restriction
is that the entire computation is carried out in a bounded
number of rounds.
Bounded-depth polynomial-size circuits cannot even compute or approximate the parity of $n$ bits, but
they can approximate real functions described by
bounded-degree polynomials and can sample approximately
according to probability distributions described by
real polynomials of bounded degree.

\begin{theorem}[Kalai and Kindler \cite{KalKin14}]
\label {thm:1}
When the noise level is constant, BosonSampling distributions are well approximated
by their low-degree Fourier--Hermite expansion.
Consequently, noisy BosonSampling can be approximated by
bounded-depth polynomial-size circuits.
\end {theorem}

It is reasonable to assume that for all proposed implementations of BosonSampling
the noise level is at least a constant, and therefore, an experimental realization
of BosonSampling represents, asymptotically, bounded-depth computation.
The next theorem shows
that
implementation of BosonSampling will actually require pushing down the noise level
to below $1/n$.

\begin{theorem}[Kalai and Kindler \cite{KalKin14}]
\label {thm:2}
When the noise level is $\omega(1/n)$, and $m \gg n^2$,
BosonSampling  is very sensitive to noise with a
vanishing correlation between the noisy distribution
and the ideal distribution.\footnote{The condition $m \gg n^2$ can
probably be removed by a more detailed analysis.}
\end {theorem}

Theorems \ref{thm:1} and \ref {thm:2} give evidence
against expectations of demonstrating
``quantum supremacy'' via BosonSampling: experimental BosonSampling
represents
an extremely low-level computation,
and there is no precedence for a ``bounded-depth machine''
or a ``bounded-depth algorithm'' that gives a practical advantage, even
for small input size,  over the full power of classical computers, not
to mention some superior powers.\footnote{While not demonstrating quantum supremacy,
I expect that BosonSampling for $n$ bosons with {\it two} modes
to be experimentally beyond reach for a few tens of Bosons.}

\begin{figure}
\centering
\includegraphics[scale=0.5]{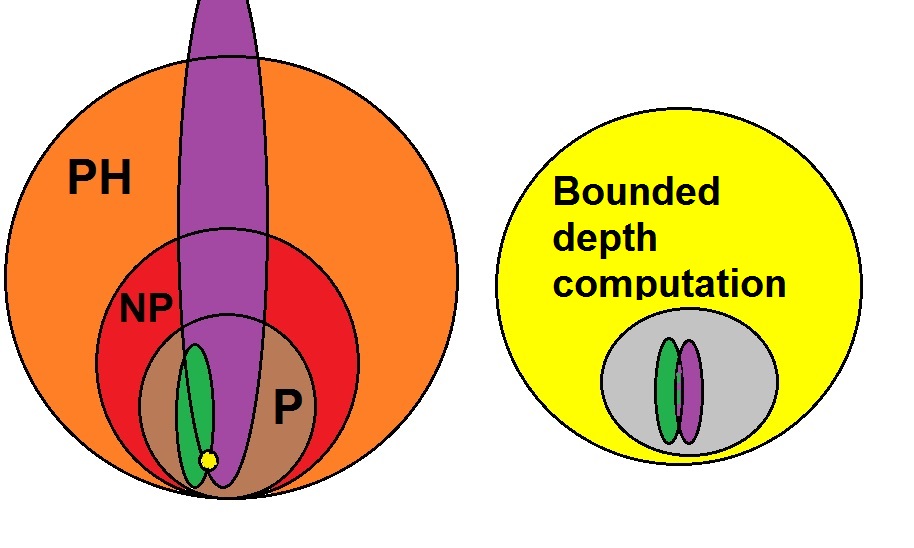}
\caption{The huge computational gap (left) between BosonSampling (purple) and FermionSampling (green)
vanishes in the noisy versions (right).}
\label{fig:1}
\end{figure}

\subsection {Bounded-degree polynomials}
The class of probability distributions that can be approximated by low-degree polynomials
represents a severe restriction below bounded-depth computation.
The description of noisy BosonSampling with low bounded-degree polynomials
is likely to extend to small noisy quantum circuits and other
similar quantum systems and this would support the
pessimistic hypothesis.
This description is relevant to
important general computational aspects of
quantum systems in nature, to which we now turn.

\subsubsection*{ Why is robust classical information  possible?}
The ability to approximate
low-degree polynomials
still supports robust {\it classical} information.
The (``Majority'') Boolean function\footnote{A Boolean function is a
function from $\{-1,1\}^n$ to $\{-1,1\}$.}
$f(x_1,x_2,\dots,x_n) = sgn (x_1 + x_2 + \cdots + x_n)$
allows for very robust bits
based on a large number of noisy bits and
admits
excellent low-degree approximations.
Quantum error-correction is also based on encoding a single qubit as a function
$f(q_1,q_2\dots,q_n)$ of many qubits, and also for quantum codes, the quality of the encoded qubit
grows with the number of qubits used for the encoding. But for quantum error-correcting codes, implementation with
bounded-degree polynomial approximations (or even with law-depth computation) is not available,
and I conjecture that no such implementation exists.
This  would support the insight that 
quantum mechanics is limiting the information one can extract
from a physical system in the absence of mechanisms
leading to robust  classical information.

%{\bf Remark: }
%One of my motivations for conducting research on quantum systems and noise came
%from
%Michel Devoret's 2002
%lecture at Yale entitled
%``The Quantum Computer: Miracle or Mirage?''
%The abstract of Devoret's lecture reflecting an optimistic view, reads:
%``Quantum mechanics is still too often viewed as limiting the information
%one can extract from a physical system.
%Recent discoveries show
%that fundamental properties of quantum mechanics could actually
%be used to perform computations that would be impossible on a
%standard `classical' computer.''
%The picture described here supports physicists' conservative intuitions on
%quantum mechanics and information.
%%%%Quantum mechanics is actually limiting the information
%%%%one can extract from a physical system. QExtracting robust information requires mechanisms leading to robust
%%%%classical information.

\subsubsection*{Why can we learn the laws of physics from experiments?}
We talked about how hard it is to compute a known function. When we need to {\it learn} an unknown function
we find ourselves in the realm of computational 
learning theory, a central area in 
the theory of computing with strong relations 
to artificial intelligence and to statistics \cite{KeVa94,ShBe14}.
Learning the parameters of a process from examples can be computationally intractable,
even if the process belongs to a low-level computational task.
Learning even a function described
by a depth-two Boolean circuit
of polynomial size does not admit an efficient algorithm. Daniely, Linial,
and Shalev-Shwartz \cite{DLS13}
showed (under certain computational complexity assumptions)
that general functions in ${\bf AC}^{\bf 0}$ (even of depth two)
cannot be efficiently learned; namely,  there is no
efficient algorithm for learning the function by
observing a small number of random examples. 
However, the approximate value of a low-degree polynomial
can efficiently be learned from examples.
This offers a theoretical explanation for our ability to understand
natural processes and the parameters defining them.

\subsubsection* {Reaching ground states}
Reaching ground states is computationally hard ({\bf NP}-hard) even for classical
systems, and for quantum systems it is even harder.  %(It represents a larger complexity class {\bf QMA}.)
So how does nature reach ground states so often? The common answer relies on two ingredients:
the first is that physical systems operate in positive temperature rather than zero temperature, and
the second is that nature often reaches meta-stable states rather than ground states.
However, these explanations are incomplete: we have good theoretical reasons to think that,
for general processes in positive temperature,
reaching meta-stable states is computationally intractable as well.
First, for general quantum or classical systems,
reaching a meta-stable state can be
just as computationally hard as reaching a ground state.
Second, one of the biggest breakthroughs in computational complexity,
the ``PCP-theorem'' (in physics disguise),
asserts that positive temperature offers no computational complexity relief for general
(classical) systems.
%In Section \ref {s:bs2}
%we proposed that rudimentary noisy quantum states
%and evolutions can be described via low-degree polynomials.
Dealing with quantum evolutions and states %that represent states
approximated by low-degree polynomials may
support the phenomenon of easily reached ground states.

\section {Predictions from the pessimistic hypothesis}

\label {s:pred}

Under the pessimistic hypothesis, universal quantum devices are
unavailable, and we need to devise a specific device
in order to implement a specific quantum evolution.
A sufficiently detailed
modeling of the device will lead to a familiar detailed Hamiltonian modeling of the quantum
process that also takes into account the environment and various forms of noise.
Our goal is different: we want to draw from the pessimistic hypothesis  predictions on noisy quantum
circuits (and,  at a later stage, on more general noisy
quantum processes) that are common to {\it all}
devices implementing the circuit (process).

The basic premises for studying noisy quantum evolutions when the specific quantum devices are not
specified are as follows: first, modeling is implicit;  namely, it is given in terms of
conditions that the noisy process must satisfy.
Second, there are systematic relations between the noise and the entire quantum
evolution and also between the target state
and the noise.

In this section we assume the pessimistic hypothesis, but we note that the
previous section proposes the following picture
in support of the pessimistic hypothesis:
evolutions and states of quantum devices in the small scale are described by low-degree polynomials.
This allows, for a larger scale, the creation of robust
classical information and computation,
but does not provide the necessary starting point for quantum fault-tolerance or
for any manifestation of
quantum supremacy.

%Some researchers regard noise as
%solely an engineering issue, and others even posit that noise
%does not have any objective meaning.
%These views raises interesting
%conceptual issues and are related to the question if
%quantum supremacy is a real phenomenon.}%\footnote{The pessimistic and optimistic
%hypotheses reflects different intuitions
%for the engineering difficulties of scaling up systems.
%The optimistic hypothesis relies on the belief that
%scaling up an engineering device based on $n$ elements
%represents polylogarithmic or polynomial-scale difficulty
%rather than exponential difficulty.
%In view of the picture drawn in Section 5,
%the pessimistic hypothesis is based on the following alternative:
%Scaling up an engineering device based on $n$ elements which perform asymptotically a task in
%a very low-level complexity class will fail well before the device
%demonstrates full classic computational powers or superior computational powers.}

\begin{figure}
\centering
\includegraphics[scale=0.3]{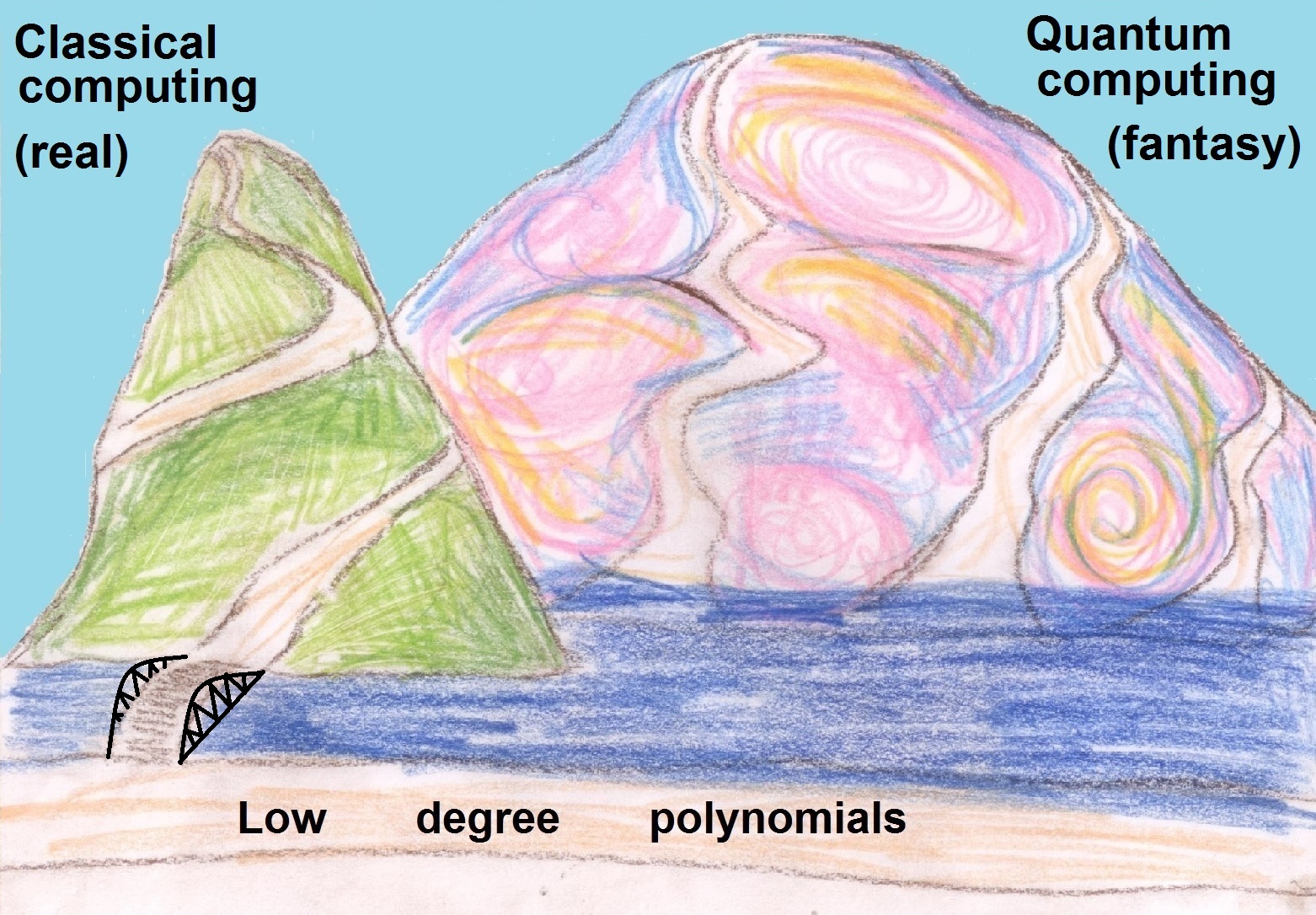}
\caption{{\bf The pessimistic hypothesis:}  Noisy quantum evolutions,
described by low-degree polynomials, allow, via the
mechanisms of averaging/repetition,
robust {\it classical} information and computation,  but do not allow reaching the
starting points for quantum supremacy and quantum fault-tolerance.
 Drawing by Neta Kalai.}

\label{fig:pes}
\end{figure}

\subsection {No quantum fault-tolerance: Its simplest manifestation}

\noindent

\subsubsection* {Entanglement and cat states}
Entanglement is a name for quantum correlation,
and it is an important feature of
quantum physics and a crucial ingredient of quantum computation.
A \emph{cat state} of the
form ${\frac{1}{\sqrt 2}}\left|00\right\rangle +{\frac{1}{\sqrt 2}} \left|11\right\rangle$
represents
the simplest form of entanglement between two qubits.
Let me elaborate: the Hilbert space $H$
representing the  states of a single qubit is two-dimensional. We denote
by $\left|0\right\rangle$
and  $\left|1\right\rangle$ the two vectors of a basis for $H$.
A pure state of a qubit is a {\em superposition}
of basis vectors of the form
$a  \left|0\right\rangle +b  \left|1\right\rangle $, where $a,b$ are complex and $|a|^2+|b|^2=1$.
Two qubits
are represented by a tensor product $H \otimes H$ and we denote
$  \left|00\right\rangle = \left|0\right\rangle \otimes \left|0\right\rangle$, and
 $  \left|11\right\rangle = \left|1\right\rangle \otimes \left|1\right\rangle$.
 A superposition of two vectors  can be thought of as a
quantum analog of a coin toss in classical probability. The superposition  ${\frac{1}{\sqrt 2}}\left|0\right\rangle +{\frac{1}{\sqrt 2}} \left|1\right\rangle$ is a quantum analog of a coin toss giving head with probability 1/2 and tail with probability 1/2, and
the superposition ${\frac{1}{\sqrt 2}}\left|00\right\rangle +{\frac{1}{\sqrt 2}} \left|11\right\rangle$
is a quantum analog of correlated
coin tosses, i.e., two heads
with probability 1/2, and two tails with probability 1/2.
The name ``cat state'' refers, of course, to Schr\"odinger's cat.

\subsubsection* {Noisy cats}

The following prediction regarding noisy entangled pairs of qubits (or ``noisy cats'')
is perhaps the simplest prediction on noisy quantum circuits under the pessimistic hypothesis.

\noindent

{\bf Prediction 1: Two-qubits behavior.}
Any implementation of quantum circuits is subject to %(depolarizing)
noise, for which errors
for a pair of entangled qubits will have substantial positive correlation.

Prediction 1, which we will refer to as the ``noisy cat prediction,''
gives a very basic difference between the optimistic and pessimistic hypotheses.
Under the optimistic hypothesis, gated qubits will manifest correlated noise, but when quantum fault-tolerance is in place,
such correlations will be diminished for most pairs of qubits. Under the pessimistic hypothesis
quantum fault-tolerance is not possible,
and without it there is no mechanism to remove correlated noise for entangled qubits.
Note that the condition on noise for a pair of entangled qubits is implicit as it depends on the
unknown process and unknown device leading to the entanglement.

\subsubsection* {Further simple manifestations of the failure of quantum fault-tolerance}

\noindent

{\bf Prediction 2: Error synchronization.} For complicated (very entangled) target
states, highly synchronized errors will occur.

Error synchronization refers to a substantial probability that a large number of qubits, much beyond the average rate of noise,
are corrupted. Under the optimistic hypothesis error synchronization is an extremely rare event.

{\bf Prediction 3: Error rate.} For complicated evolutions, and for evolutions approximating complicated states,
the error rate, in terms of qubit errors, scales up linearly with the number
of qubits.

%\footnote {We note that it
%is crucial for the pessimistic hypothesis
%that error-rate
%in terms of qubit-errors
%for implementations
%of universal (or complicated) quantum circuits scales up with the number of qubits.
%If we assume arbitrary form of correlation but small error-rate (in terms of qubit errors)
%then log-depth quantum computation (hence Shor's factoring) is still possible \cite {Kal06}.
%There are measures of rate (in terms of some operators norm) used for Hamiltonian noise
%models \cite {TerBur05,Pre13}, where small rate suffices for full-fledged quantum fault tolerance.
%Those results and those by by Kalai and Kuperberg
%\cite {KalKup14} on contagious noise models
%show that when you force error-rate to be small, some
%general forms of correlation still allow quantum fault-tolerance.}

The three predictions 1--3 are related.
Under natural assumptions,
the noisy cat prediction implies error synchronization for quantum states of
the kind involved in quantum error-correction and quantum algorithms. Roughly speaking, the noisy cat prediction
implies positive correlation between errors for every pair of qubits, and this implies a substantial probability
for the event that a large fraction of qubits (well above the average rate of errors)
will be corrupted at the same computer cycle. Error synchronization
also implies, again under some natural assumptions,
that error rate in terms of qubit errors is at least linear in the number
of qubits.
Thus, the pessimistic hypothesis itself can be justified from
the noisy cat prediction together with natural assumptions on the rate
of noise. Moreover, this also explains the wide-gap dichotomy in terms of qubit errors.

The optimistic hypothesis allows creating via quantum error-correction very stable ``logical'' qubits based
on stable raw physical qubits.

{\bf Prediction 4: No logical qubits.} Logical qubits cannot be
substantially more stable than the raw qubits used to construct them.

\subsection {A more formal description of the noisy cat condition}
\label {s:fd}

Given two qubits $q_1,q_2$ in
pure joint state $\rho$ the entropy of one of the qubits is a standard measure of entanglement
that we denote by $Ent(\rho:q_1,q_2)$.
We will consider depolarizing noise described by a $2 \times 2$ matrix $(p_{i,j})$
describing the probabilities of none, only the first, only the second, and
both qubits being corrupted. Let $E_i, i=1,2$ be the event that the $i$th qubit was corrupted and
let $r_i$ be the probability of $E_i$ and $cor(E_1,E_2)$ be the
correlation between the events $E_1$ and $E_2$.

The noisy cat prediction asserts that any realization of a quantum circuit that
approximates the pure state $\rho$ is subject to depolarizing noise with
$$cor (E_1,E_2) \ge K(r_1,r_2) \cdot Ent(\rho:q_1,q_2).$$
Here, $K(x,y)$ is a function of $x$ and $y$ so that
$K(x,y)/ \min(x,y)^2 \gg 1$
when $x$ and $y$ are
positive and small.

A stronger form of the prediction applies to the {\it emergent entanglement}
of a pair of qubits,
namely, entanglement after some other qubits are measured (separately).
The emergent entanglement of pairs of qubits is large both for quantum error-correcting codes needed for
quantum fault-tolerance and quantum algorithms. The strong form of the prediction will
imply substantial correlation of the noise between every pair of qubits.
This implies error synchronization for quantum states of
the kind involved in quantum error-correction and quantum algorithms.
Assuming further that rate in terms of {\it trace distance} is constant for short time intervals
implies that error rate in terms of qubit errors is
at least linear in the number of qubits. This, in brief, is the
reason for the wide-gap dichotomy
between the optimistic and pessimistic hypotheses in terms of qubit errors.
It is natural to assume that noise in terms of trace distance, (namely, bounded variation distance
between probability distributions)
is, for short time intervals, constant, because trace distance
is invariant under unitary
transformations. For more details see \cite {Kal11}.

%Consider now depolarizing noise based on a probability
%distribution $D$ on $\{0,1\}^n$.
%First draw a random vector $w$ according to $D$ and associated to
%every $\{0,1\}$ vectors $w$ of length $n$ and associate to $w$ the event
%that the qubit $k$ will be damaged iff $w_k=1$.
%%%%The noise operation will be described in terms of a probability distribution on $\{0,1\}^n$.
%The rate of noise $r_k(E)$ for the $k$th qubit is the probability that $w_k=1$
%and the correlation $cor_{i,j}(E)$ is the correlation between the events $w_i=1$ and $w_j=1$.
%(The reader is referred to \cite {Kal,Kal} for more general treatement.)

%Without getting into the full technical aspects of Predictions 1 and 2 we
%will make some additional comments.

A few comments: first, to deal with noise it is very
important to understand general sources and forms of noise,
but for showing that noise cannot be dealt with,
%when we assume that the
%overall rate of noise is small,
we can safely assume that depolarizing noise is present and
restrict the discussion to depolarizing noise.
Generally speaking, in this section
we present predictions about noise when the system approximates well
some ideal noiseless quantum state or evolution. Additional forms of
noise may also be present.
%We still need to
%assume that the overall rate of noise is small.
We don't expect that (in practice)
additional forms of noise will heal the damaging effects of our predicted noise.
(This is theoretically possible \cite {BGH13}.)
Second, regardless of the possibility of quantum fault-tolerance we can expect our
predictions to apply to any implementation of small quantum computers. Third,
we note that arbitrary forms of correlation with a small error rate (in terms of qubit errors)
still likely support log-depth quantum computation (hence Shor's factoring)  \cite {Kal06}.
However, as discussed above, under natural assumptions,
strong forms of correlation imply increased error rate in terms of qubit errors and vice versa.

\subsection {No quantum fault-tolerance: its most general manifestation}
%\subsection {No quantum fault-tolerance: Its most general manifestation\protect\footnote
%{This section is more technical and assumes more background on quantum information. }}
\label {s:gm}

We can go to the other extreme and try to examine consequences of the pessimistic hypothesis
for the most general quantum evolutions.
We start with a prediction related to the discussion in Section 5.

{\bf Prediction 5: Bounded-depth and bounded-degree approximations.}
Quantum states achievable by
any implementation of quantum circuits are limited by
bounded-depth polynomial-size quantum computation.
Even stronger: low-entropy quantum states in nature
admit approximations by bounded-degree polynomials.

The next items go beyond the quantum circuit model
and do not assume that the Hilbert space for our quantum evolution has a tensor product structure.

{\bf Prediction 6: Time smoothing.}
Quantum evolutions are subject to noise with a substantial correlation with time-smoothed evolutions.

Time-smoothed evolutions
form an interesting restricted class of noisy quantum evolutions
aimed for modeling evolutions under the pessimistic hypothesis when fault-tolerance is
unavailable to suppress noise propagation.
The basic example for time-smoothing is the following:
start with an ideal quantum evolution
given by a sequence of $T$ unitary operators,
where $ U_t$ denotes the unitary
operator for the $t$-th step, $t=1,2,\dots T$.
For $s<t$ we denote $ U_{s,t} = \prod_{i=s}^{t-1}U_i$ and let $U_{s,s}=I$ and $U_{t,s}=U^{-1}_{s,t}.$
The next step is to add noise in a completely standard way: consider a noise operation $ E_t$ for
the $t$-th step. We can think about the case where the unitary evolution is a quantum computing process
and $E_t$ represents a depolarizing noise with a fixed
rate acting independently on the qubits.
And finally, replace $E_t$ with a new noise operation $E'_t$ defined as the average
%of $U_{s,t} E_s U^{-1}_{s,t}$,
%over $s=1,2,\dots,T$.

\begin {equation}
\label {e:smooth}
 E'_t =
\frac{1}{T}
\cdot
\sum_{s=1}^T  U_{s,t} E_s U^{-1}_{s,t}.
\end {equation}

{\bf Prediction 7: Rate.}  For a noisy quantum system a lower bound for the rate of noise in a time interval
is a measure of non-commutativity for the projections in the algebra of unitary operators
in that interval.

Predictions 6 and 7 are implicit and describe systematic relations between
the noise and the evolution.
We expect that time-smoothing will suppress
high terms for some Fourier-like expansion, thus relating Predictions 6 and 5.
We also note that Prediction 7 resembles the picture of the ``unsharpness principle''
from symplectic geometry and quantization \cite{Pol14}.
%\footnote{L. Polterovich,  Symplectic geometry of quantum
%noise, {\it Comm Math. Phys} 327 (2014), 481--519,
%arXiv:1206.3707.}

\subsection {Locality, space, and time}
\label {s:locality}

The decision between the optimistic and pessimistic hypotheses,
is, to a large extent, a question
about modeling locality in quantum physics.
Modeling natural quantum evolutions by quantum computers represents the
important physical principle of ``locality'':
quantum interactions are limited to a few particles. The quantum circuit model enforces
local rules on quantum evolutions and still allows the creation of very
non-local quantum states. This remains true for noisy
quantum circuits under the optimistic hypothesis.
The pessimistic hypothesis suggests that quantum supremacy is an artifact of incorrect modeling of locality.
We expect modeling based on the pessimistic hypothesis,
which relates the laws of the ``noise'' to
the laws of the ``signal,'' to force a strong form of locality
for both.

We can even propose that spacetime itself emerges
from the absence of quantum fault-tolerance.
It is a familiar idea that since (noiseless)
quantum systems are time-reversible,
time emerges from quantum noise (decoherence).
However, also in the presence of noise, with quantum fault-tolerance, every quantum
evolution that can experimentally be created
can be time-reversed and, in fact, we can time-permute
the sequence of unitary operators describing the evolution in an arbitrary way.
It is therefore both quantum noise and the absence of quantum fault-tolerance that
enable an arrow of time.

Next, we note that with quantum computers one can emulate
a quantum evolution on an arbitrary geometry. For example, a complicated  quantum
evolution representing the dynamics of
a four-dimensional lattice model could be emulated on a one-dimensional chain of qubits.
This would be vastly different from today's
experimental quantum physics, and it is also
in tension with insights from  physics, where witnessing
different geometries supporting the same physics is rare and important.
Since a universal quantum computer
allows the breaking of the connection between
physics and geometry, it is noise and the absence of
quantum fault-tolerance that distinguish physical processes based on different geometries
and enable geometry to emerge from physics.

\subsection {Classical simulations of quantum systems}

\noindent

{\bf Prediction 8: Classical simulations of quantum processes.} Computations in quantum physics can,
in principle, be simulated efficiently on a digital computer.

This
bold prediction from the pessimistic hypothesis
could
lead to specific models and computational tools.
There are some caveats: heavy computations may be required (1) for quantum processes that are not
realistic to start with, (2) for a model in quantum physics
representing a physical process that depends on
many more parameters than those represented by the input size,
(3) for simulating processes that require knowing internal parameters of the process that
are not available to us (but are available to nature),
and (4) when
we simply do not know the correct model or relevant computational tool.

\section {Additional predictions from the pessimistic hypothesis}
%: quantum refrigerators, fluctuations, symmetry, time, geometry, and teleportation}

\label {s:pre2}
%\noindent
We describe here a few further predictions from the pessimistic hypothesis.
There are classes of quantum states that require
deep (namely, of large depth) quantum computing and are thus unattainable under the
pessimistic hypothesis. Since mixed states have multiple representations
in terms of pure states, within a symmetry class of quantum states (or
a class described by other terms), it is possible that low-entropy states will
not be supported by low-degree polynomials and will thus be infeasible,
while higher-entropy
states in the class will admit low-depth/low-degree description and will
actually be feasible. This leads to:

{\bf Prediction 9: Cooling.} Within a symmetry class of quantum states (or
for classes of states defined in a different way), the bounded-depth/low-degree polynomial
requirement provides an absolute lower bound for cooling.

Of course, reaching low-temperature states in a certain class of quantum states may reflect
a harder engineering task
under both hypotheses. Under the pessimistic hypothesis, however, we may actually witness some
threshold (depending on the class) that we cannot cross as the engineering difficulty explodes.
%Again, the hardness of computing Ramsey number is a good role model
This remark applies to a few of the other predictions below.
(Here, the explosion of difficulty of computing Ramsey numbers is a good ``role model.'')

We outline an important special case. Anyonic
states \cite {MR91,Kit03} are of special interest,
both on their own and as a potential avenue for quantum computing.

{\bf Prediction 10: Anyons.} Stable anyonic qubits %, and low-entropy anyonic states
cannot be constructed.

Next, under the pessimistic hypothesis the noisy process
leading to a quantum state with a certain symmetry will
introduce noise obeying the same symmetry. (Of course,
other forms of noise may also be present.)

{\bf Prediction 11: Symmetry.} Noisy quantum states and
evolutions are subject to noise that
respects their symmetries.

An interesting
example is that of Bose--Einstein condensation. For a
Bose--Einstein state on a bunch of atoms, one type of noise corresponds to independent
noise  for the individual atoms. (This type of noise is similar to
standard noise for quantum circuits.) Another type of noise represents
fluctuations of the collective Bose--Einstein state itself. This is the noise
that respects the internal symmetries of the state and we expect
that under the pessimistic hypothesis such a form of noise must always be present.
%(The second type is the one represented by my smoothed Lindblad.)

Our next prediction challenges one of the consequences
of the general Hamiltonian models allowing
quantum fault-tolerance \cite {TerBur05,AGP05,Pre13}. These models allow some
noise correlation over time and space but they are
characterized by the fact that the error fluctuations are sub-Gaussian. Namely,
when there are $N$ qubits the standard deviation
for the number of qubit errors behaves like $\sqrt N$
and the probability of more than $t \sqrt N$ errors decays as it does for Gaussian distribution.
These properties are not necessary for quantum fault-tolerance but
they are shared by the rather general Hamiltonian models for noisy
quantum computers that allow quantum fault-tolerance.

{\bf Prediction 12: Fluctuation.}
Fluctuations in the rate of noise for interacting $N$-element systems
(even in cases where interactions
are weak and unintended) scale like $N$ and not like $\sqrt N$.

Our prediction about fluctuation of noise for interacting systems
can be tested in a variety of places. For example, we can consider
a single superconducting qubit as a quantum
device based on a large number of microscopic elements and study
how stable its instabilities are. Prediction 4.4 can also be tested
on digital memories (where interactions are unintended). Systems for
highly precise physical clocks are
characterized by having a  huge number $N$ of elements with extremely
weak interactions. We still expect (and this may be supported
by current knowledge)
that
in addition to $\sqrt N$-fluctuations there will be also
some $\epsilon N$-fluctuations. The relation between the
level of interaction and $\epsilon$ can be useful for making quantitative versions of 
our predictions on correlated noise for systems with interaction.
Another well-studied issue that might be relevant
here is the statistical behavior of decay-time of
particles and various other quantum systems.

{\bf Prediction 13: Teleportation.} Teleportation of complicated quantum
states is not possible.

Teleportation is an important, well-studied, and experimentally tested quantum physics
phenomenon. Under the pessimistic hypothesis there are quantum states described
efficiently by quantum circuits that are beyond reach. For complicated quantum states that
are realistic, since teleportation itself involves additional noise and would require quantum
fault-tolerance, teleportation need not be possible. With the absence of
quantum fault-tolerance,
quantum teleportation of complex %macroscopic
quantum systems may well be impossible.

Again, an example based on non-interacting bosons is in order.
Consider a photonic implementation of BosonSampling.
We expect that already for fairly low values of $n$ we
will not be able  to reach with good accuracy
BosonSampling states based on random Gaussian $n \times 2n$
matrix for  $n$ photons and $2n$ modes.
%(Here the analogy with computing  Ramsey numbers, is useful.)
Let's suppose that the threshold will be $n=8$.
The threshold may well go down to $n=4$ or $n=5$ if we pose a more difficult task
of achieving such a goal and then teleporting our $n$ photons to $n$ different
locations  100 miles apart.

{\bf Prediction 14: Reversing the arrow of time.} There are
quantum evolutions that can be demonstrated but cannot be time-reversed.

Under the pessimistic hypothesis there are quantum
evolutions (described, say, by quantum circuits) that
cannot be realized (approximately). We can expect that the class
of realistic noisy quantum evolutions is not invariant under
time-reversing, and that there are easy-to-implement evolutions whose
time-reversed versions are infeasible.
(Of course, we do not restrict ourselves to actual physical
implementations of quantum circuits via qubits and gates.) Thus,
it may well be the case that we cannot, in principle, turn an
omelet into an egg. (But simpler examples
would certainly be desirable.) By a similar token:

\begin{figure}
\centering
\includegraphics[scale=0.4]{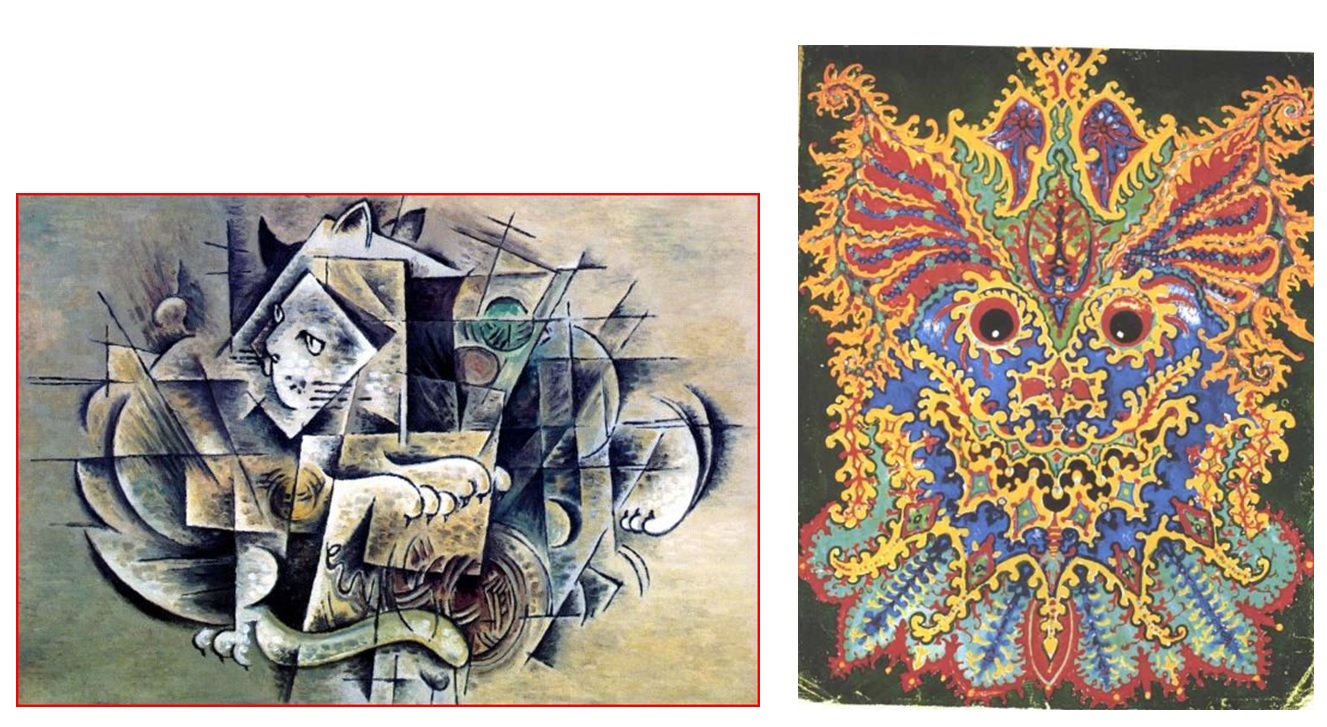}
\caption{ Cats with unusual geometry in paintings by Gianantonio 
Muratori (in Braque's style, left) and by Louis Wain (right) } 
\label{fig:cat}
\end{figure}

{\bf Prediction 15: Geometry.} Quantum states and evolutions reveal
some information on the geometry of (all) their physical realizations.

%Like with the prediction in Section \ref{s:pre}

{\bf Prediction 16: Superposition.} There are pairs of
quantum states/evolutions that can be created
separately but cannot be superposed.
%\footnote{Moty Heilborn described (private communication) such a behavior
%for experiments of certain anyonic systems.}

Quantum noise and the absence of quantum fault-tolerance leads  also to:

{\bf Prediction 17: Predictions.} Complex quantum systems cannot be predicted.

Consider again an experiment aimed at approximating the quantum
state of  BosonSampling based on a random Gaussian matrix,
with $n$ bosons and $2n$ modes.
In view of \cite {KalKin14} (see, Section \ref{s:bs2}), we expect that the engineering effort required 
% for a few tens of bosons
%we cannot expect 
for a noise level below $1/n$, will explodes already for a small value of $n$. (Say, $n=8$.) 
On the other hand, for somewhat larger values of $n$ (say, $n=12$) 
when the noise level is above $1/n$,
%(but not very large so the experiment
%carrys some information),
the experimental outcomes will not be robust. Noise sensitivity does not
allow robust experimental outcomes because of
the dependence of the state on an exponential number of parameters describing the actual
noise \cite{KalKin14}[Sec. B.4]. Therefore, it will not be possible (in principle)
to predict the outcome of the experiment,
even in terms of a probability distribution.\footnote{A reader may ask whether
the outcomes of such an experiment
represent superior computational power. The answer is negative since what the
experiment actually represents is a computational process
that is computationally simple but dependent on
superexponential size input; \cite {KalKin14}. We have to distinguish between computational hardness in terms of
the running time of an algorithm as a function of the input size,
and in terms of the input size.}$^{,}$\footnote{We note that inability to predict implies neither a
computational advantage nor a disadvantage. We cannot predict the 
next move of a human chess master (who exercises
his free will and best judgement in making moves), but we can perfectly 
predict moves of computer chess programs. On the other hand,
it would not be difficult to replace a chess-playing program by a comparable one that is unpredictable.}

As before, it will be desirable to find concrete, formal, and
quantitative versions of all these predictions. We note
also that it can be an interesting mathematical challenge to relate the different
predictions based on the pessimistic hypothesis. One example would be to prove that noise preserves symmetry for
noise described by time-smoothing.

\subsection {Predictions for a living cat}

Following the tradition of using cats
for quantum thought experiments,
consider an ordinary living cat. (An ordinary cat,
not a Schr\"odinger cat.) All the difficulties predicted
based on the pessimistic hypothesis for
a handful of non-interacting photons positioned in interesting quantum states
are expected to apply to the cat as well.
Under  the pessimistic hypothesis and the
in-principle absence of quantum fault-tolerance,
it will be impossible
%we will not be able %, in principle,
to teleport the cat,
it will be impossible
%we will not be able %, in principle,
to reverse the life-evolution of the cat,
it will not be possible
%we will not be able, %in principle,
 to implement the cat at a very low temperature, or on a device with very different geometry,
it will be impossible
%we will not be able, %in principle,
to superpose the life-evolutions of two distinct cats,
%there will be specific lower bounds for cooling a living cat,
and, finally, we predict that, even if we place the cat in an isolated
and monitored environment,
the life-evolution of this cat cannot %, in principle,
be predicted.%\footnote {Of course, these predictions about
%cats may seem plausible (or even valid, in principle), even to many of
%those who are optimistic regarding quantum supremacy and quantum fault-tolerance.}
%%%However,
%%%the pessimistic hypothesis can be proposed as a clear %conceptual %physical
%%%principle to support it.}
%\footnote{Also, a spherical cat will have a spherical meow.}

\section {Discussion}

\subsection {Noise and scalability}

\subsubsection *{The emerging picture}

Let me start by summarizing  the picture drawn in Sections 5, 6, and 7.
Standard noise models for evolutions and states of quantum devices in the small scale
%can be
%modelled by the standard noise models and this
lead to  a description by low-degree polynomials
that manifests very low computational power that does
not allow quantum fault-tolerance and quantum supremacy to emerge.  This behavior in the small scale
 allows, for larger scales, the creation of robust classical information and computation (Section 5).
The inability to reach the starting point for quantum fault-tolerance in the small scale
has far-reaching consequences for the behavior of noisy quantum systems
in the medium and large scale. In particular, it leads to novel implicit
ways for modeling noise (Sections 6) that express the additional property of
``no quantum fault-tolerance.''
%It implies that devices for universal quantum circuits
%cannot be realized

\subsubsection *{The nature of noise}

Some researchers regard noise 
solely as an engineering issue, and others even posit that noise
does not have any objective meaning.
I tend to disagree but, in any case, these views raise interesting
conceptual issues and are related to the question of whether
quantum supremacy is a real phenomenon.
%A related claim is that any principle regarding noise should
%be driven from the Schr\"odinger equation itself.

\subsubsection *{The gap in intuitions regarding scaling}

The pessimistic and optimistic
hypotheses reflect different intuitions
of the %engineering
difficulties of scaling up systems.
The optimistic hypothesis relies on the belief that
scaling up an engineering device based on $n$ elements
represents polylogarithmic or polynomial-scale difficulty
rather than exponential difficulty.
In view of the picture drawn in Section 5,
the pessimistic hypothesis is based on the following alternative:
scaling up an engineering device based on $n$ elements that perform asymptotically a task in
a very low-level complexity class will fail well before the device
demonstrates full classic computational powers or superior computational powers. Note that the pessimistic
intuition about scaling is supported by computational complexity considerations 
applied (rather unusually) to small systems.

\subsubsection* { Modeling by quantum circuits}
Under the pessimistic hypothesis, universal quantum circuits are
beyond reach and they cannot be achieved
even for a small number of qubits. But quantum circuits remain a
powerful framework and model for quantum evolutions.  Abstract quantum circuits are general enough
to model processes in quantum physics (in fact, vastly more general),
but note that we cannot take for granted that any
realistic quantum process can be realized by a realistic implementation of a quantum circuit.

\subsubsection* {Computers and circuits} There are two slightly different ways to interpret ``realizing universal 
quantum circuits'' referred to in our hypotheses. Let us consider quantum circuits based on a fixed 
universal set of quantum gates. The simplest 
interpretation that we use in the paper is that we seek devices which realize arbitrary 
such circuits and we allow a special device for each 
circuit. The optimistic hypothesis 
asserts that with feasible engineering efforts 
every such circuit could be realized with a bounded and small error rate per qubit per computation step. 
The pessimistic hypothesis asserts that even for specific circuits needed for quantum algorithms, 
the error rate would scale up and the engineering effort required for keeping it small would thus explode.     
%As a matter of fact, 
The optimistic view is actually slightly stronger: it asserts that one can build a 
universal controlled device,  a quantum computer,  
that would allow to implement every quantum computational process, just like 
a digital computer can implement every Boolean computation.

\subsubsection* {How does the future evolution affect the present noise?}

Under the pessimistic hypothesis,
 there is a systematic relation
between the law for the noise at a given
time and the entire evolution, including the future evolution.
This is demonstrated by our smoothed formula (\ref{e:smooth}).
We can ask how can the current noise (or risk)
depend on the future evolution.
The answer is that it is not that the evolution in the future
causes the behavior of the noise in the past, but
rather that the noise in the past
leads to constraints on possible %feasible
evolutions in the future.
Such dependence occurs also in classical systems. Without
refueling capabilities, and without a very
detailed description of the spacecraft,
calculating the risk of space missions at take-off will strongly
depend on the details of the full mission.
(Such dependence can largely be be eliminated with refueling
capabilities.) %A principle of no quantum fault-tolerance
The pessimistic hypothesis  implies that in
the quantum setting such dependence cannot be eliminated.

Here is another example: suppose you are told that you need to undergo an operation
at the age of fifty to avoid serious health problems in the following decade.
%you will live to the age of
%eighty with probability 90\% if you will have an
%operation at the age of fifty
%with  mortality probability of 10\%.
This makes the risk at age fifty, conditioned on living to eighty, higher.
Of course, in this scenario, it is not that living to eighty raises the risk at fifty,
but rather that not taking the risky alternative at fifty makes it impossible to live to eighty.

%\subsubsection *{The puzzle and the debate}

%While I cannot give full review of views regarding the debate on quantum computers and quantum fault tolerance
%since the mid 90s let me cite some references.

%\subsubsection *{The debate}

\subsection {Some further connections with physics, mathematics, and computation.}

\subsubsection* {Quantum simulators, and proposals for quantum systems
that cannot be simulated classically.}

From time to time there are claims regarding quantum supremacy being
manifested in some special-purpose
quantum devices and, in particular, quantum simulators that  simulate  some quantum
systems via quantum devices of a different nature. One such claim is outlined by
a recent blog comment (\href{http://www.scottaaronson.com/blog/?p=2448#comment-814503}{Shtetl Optimized, September 2015}) by
%Mathias
Troyer: % \cite {Tro:so}:
``The most convincing work so far might be \cite {TCF+12}
%http://arxiv.org/abs/1101.2659,
which looks at the dynamics of a correlated Bose gas.
There the quantum simulation agrees with state of the art
classical methods as long as they work, but the
quantum simulation reaches longer times. The reason is a growing
entanglement entropy which at some point causes the classical simulations to become unreliable.
This is however one of the first demonstrations
of a quantum simulator providing results that we don't have a classical algorithm for.''

Like other supreme powers, quantum supremacy is appealing and has notable explanatory
capability.  I expect, however,  that, in this case, the phenomenon of ``growing entanglement entropy,''
which causes classical simulations to become unreliable,
amounts to the decay of some high-degree coefficients in a Fourier-like expansion, similar to
the situation of noisy BosonSampling
studied in \cite {KalKin14}, and that the classical simulator
can be replaced by a better classical simulator representing
low-level classical complexity for growing entropy.

%\section {Absence of quantum fault-tolerance and theoretical physics}
%\label {s:ph+ph}

\subsubsection *{Thermodynamics and other areas of physics}

Properties of noise and the nature of approximations in quantum physics are very important in many
areas of theoretical and experimental physics.
Absence of quantum fault-tolerance seems especially
relevant to the interface of thermodynamics and quantum physics.
For example, the relations between
the ``signal'' and ``noise'' in noise-modeling and predictions
under the pessimistic hypothesis
look similar to an important  hypothesis/rule of
Onsager %\cite {Ons}
from classical thermodynamics. % (and may be relevant to the
%controversy regarding its ``quantum counterpart.'').
%It would be interesting to explore further such a connection.
We note that Alicki and several coauthors have over the years studied
relations between quantum error-correction and thermodynamics (see, e.g.,\cite {AH06}).

We briefly mentioned in
Section \ref {s:r2d}
the relevance of quantum computing to computations in
quantum field theory. ``Noise'' may well refer to
the familiar phenomenon  that for some scale, computations
based on one theory (say QED computations) need to be corrected
because of the effect of another theory
(say, effects coming from the weak force).
Our discussion here suggests that there
could be systematic general rules for such corrections
that have a bearing both on practical computations and
on computational complexity issues.

%Exotic new phases of matter that cannot be
%described by ``shallow'' quantum processes draw
%the attentions of researchers in condensed matter physics.

%A related (welcomed) general phenomenon supported by our prediction is that
%for the study of physical theories

\subsubsection* {Locality and entanglement}

The pessimistic hypothesis
excludes the ability to create highly entangled states {\it from local
operations} (quantum computation processes).
It does not exclude the possibility of ``background'' entangled systems
that do not interact
with our ``local'' physics.
In fact, the same Hilbert space may admit different tensor product representations and hence
different local structures, so that very mundane states for the one are highly entangled
for the other. This possibility may represent real physical phenomena
both under the pessimistic and optimistic hypotheses.
(Of course, there could be ``background'' states
that cannot be described by any local structure.)

There have been recent attempts
to apply quantum information ideas and entanglement in particular to the study of
quantum gravity and other basic topics in theoretical physics. Generally speaking,
the pessimistic hypothesis is not in conflict with such ideas, and it may lead to
interesting insights into them.

\subsubsection *{Classical physics, symplectic geometry, quantization,
quantum noise, and the unsharpness principle}

The unsharpness principle is a property of noisy quantum systems that
can be proved for certain quantization of symplectic spaces.
This is studied by Polterovich in \cite{Pol14} who relies on deep notions and results from symplectic geometry
and follows, on the quantum side,
some earlier works by Ozawa \cite {Oza04}, and by Busch, Heinonen, and Lahti \cite {BHL04}.
Here the notion of noise is different. The crucial distinction is
%The crucial distinction is
between general positive operator-valued measures (POVMs) and vonNeumann observables,
which are special cases of POVMs
(also known as projector-valued POVMs). The
unsharpness principle asserts that certain noisy quantum evolutions described by POVMs must
be ``far'' from vonNeumann observables.
The amount of unsharpness is bounded below by some non-commutativity measure.
(This resembles our Prediction 7.) We note that the unsharpness principle depends on some
notion of locality: it applies
to systems based on ``small'' (displaceable) sets, where a set $X$ is displaceable
if there is a Hamiltonian diffeomorphism of the entire underlying Hamiltonian manifold
so that the image of $X$ is disjoint from $X$.
It will be interesting to pursue further mathematical relations between
the unsharpness principle, smoothed evolutions, and other issues related
to quantum fault-tolerance.
%There is a notable difference: the unsharpness principle talks about noisy measurements
%and not about noisy approximations to unitary evolutions or
%to pure states.

%One interesting feature of the quantization principle is that it applies for small
%coverings
%rate There, %

\subsubsection *{The black hole firewall information paradox}

According to the classical theory of black holes an
object (which can be a photon, and  is usually referred to as ``Alice'')
can pass through the event horizon of the black hole, never to return. 
But when you add
quantum mechanics considerations (see, e.g., \cite {Har14,AMPS12}),
you are driven to the conclusion that
the event horizon itself
 represents singularity -- the interior of the black hole does not exist.
In other words,  Alice will burn up passing through it.
%This sharp disagreement %between the
% classical theory
 %and  quantum theory
 The sharply different view arising from QM is based
on the fact that the same particle cannot be in entanglement with two different particles. It was raised in a work by 
Almheiri, Marolf, Polchinski, and Sully \cite {AMPS12}. 
(There are other forms of the paradox. For example, there is an argument that Alice
will eventually evaporate and all its
quantum information will be lost, in contrast to the reversibility of quantum mechanics.
Of course, since we lack a detailed theory for
quantum gravity, there is something tentative/philosophical about the paradox and its proposed solutions.)
It will be interesting to examine the relevance of the
pessimistic hypothesis and absence of quantum fault tolerance to the paradox. Specifically,
%the noisy-cat prediction regarding noisy entanglements to the paradox, and, in particular,
it will be interesting to study if cosmological
reasoning related to the black hole firewall paradox leads to concrete estimates
of the constants for Predictions 1 and 4. %in the noisy-cat prediction.
Hayden and Harlow \cite {HaHa13} studied connections with quantum computation and
argued that implementing 
the thought
experiment that demonstrates the unexpected singularity
at the event horizon is computationally intractable. (More precisely, verifying the expected outcome 
of the thought-experiment is intractable.) 
Maldacena and
Susskind \cite {MaSu13} offered a solution (based on what they call
the {\bf ER=EPR}  principle). According to Susskind (private communication),
for firewalls to arise requires sufficient quantum-computational power,
rather than ``ordinary interaction with the environment.''
%and
% it is important to make the distinction between
%%%%A crucial ingredient of this attempt is to assume
%%%%that Alice have
%the kind of interactions with the environment
%witnessed by quantum computers and ``ordinary interaction with the environment.''
%do not arise if Alice has only ``ordinary interaction with the environment.'' %namely the kind of interactions
%with the environment
%This, of course, can be taken for
%granted under the pessimistic hypothesis.
The pessimistic hypothesis provides a  way to define
``ordinary interaction with the environment,'' and it asserts that no
extraordinary interactions are at all possible.

\subsubsection*{More on permanents: From Polya to Barvinok}

The huge computational gap between computing
determinants and permanents makes an
early appearance with the following problem proposed by
Polya and solved by Szeg\"o in 1913  (\cite{Min78}, Ch. 1.4):
``Show that there is no linear transformation $T$ on the ($n^2$-dimensional)
space of $n\times n$ matrices, $n\ge 3$,
such that %the permanent of a matrix $M$ equals the determinant of $T(M)$.
$per (M)= det (T(A))$.''

Approximating permanents of general real or
complex matrices is {\bf \#P}-hard.
But while approximating the value of  permanents is hard in general, it is sometimes easy.
An important result by Jerrum,
Sinclair, and Vigoda \cite {JSV04}
asserts that permanents of positive real matrices can be approximated in polynomial type up to a
multiplicative factor $1+\epsilon$, for every $\epsilon>0$. Another important work on the 
computational complexity of approximating permanents is by Gurvits \cite{Gur05}.

Barvinok \cite {Bar16} recently demonstrated
remarkable results on approximating permanents. He showed that under certain constraints on the matrix's
entries, approximating the value of the permanent of an $n$ by $n$ matrix
admits a quasi-polynomial time algorithm and, moreover,
the value of the permanent can be well approximated by low- (logarithmic in $n$) degree polynomials.
One example of such a constraint is when all entries
are of the form $x+yi$, where $x \in [\delta,1]$, $\delta>0$, and $|y| \le 0.5\delta^3$.
Barvinok proved that if the permanent does not vanish in a certain region in the space of
matrices then the value of the permanent is well
approximated by low-degree polynomials well inside the region.
It will be interesting to check if, for Barvinok's good regions,
BosonSampling is stable under noise
and is practically feasible. Of special interest is the class of real matrices
with entries between $\delta$ and 1.

\subsubsection*{FourierSampling and more on BosonSampling}

Given a Boolean function $f$ that can
be computed efficiently, a quantum computer can sample
according to the Fourier coefficients of $f$ \cite {Sim94}. (A similar statement
applies to functions defined on the integers that can be efficiently
computed in terms of the number of digits.)
This computational task, called FourierSampling, is crucial for
many quantum algorithms, including Shor's factoring.
Similar computational results as mentioned for BosonSampling apply
to FourierSampling, and approximately demonstrating
FourierSampling for forty qubits or so
is also regarded as a quick experimental
path toward demonstrating quantum supremacy.

It is worth mentioning that the computational difficulty in demonstrating
BosonSampling (and FourierSampling) goes even further than what we stated in Section \ref {s:bs}.
A classical (randomized) computer equipped with a subroutine that can perform an arbitrary task in
the entire polynomial hierarchy cannot perform BosonSampling unless the polynomial hierarchy itself
collapses \cite {AaAr13}.  It is even a plausible conjecture \cite {KalKin14}
that a classical (randomized) computer equipped
with a subroutine that can perform an arbitrary task in
{\bf BQP} (or even {\bf QMA}) cannot perform BosonSampling (and FourierSampling). If true,
this would demonstrate
a computational gap between quantum decision problems and quantum sampling problems.

%\subsubsection *{smoothing}

%\subsubsection *{Locality, entanglement, quantum gravity and the emergence of  time-space}

%\subsubsection *{Special relativity and quantum field theory}

\subsubsection *{Computational power of other classes of evolutions: Navier--Stokes computation}
Quantum computing is an attempt to explore the computational power supported by quantum
evolutions and quantum physics. One may well study computational aspects of
other classes of evolutions. (See, e.g., \cite {Yao03}.)
Tao conjectures \cite {Tao15} that  systems described by
three-dimensional Navier--Stokes equations support fault-tolerance, and
universal classical computation. This conjecture would imply
finite-time blowup for those equations for certain initial conditions
and thus disprove a major open problem in mathematics. Our discussion
suggests various ways to express the negation of Tao's conjecture,
either as a mathematical alternative to Tao's proposal or as a
physical condition for  ``realistic'' Navier--Stokes evolutions
if Tao's conjecture holds.

For demonstrating  a property of ``no computation'' for a systems like
those based on the 3D Navier--Stokes equation,
we can try to derive or impose a ``bounded degree'' description
for states and evolutions described by the system. (We will need also to study whether 
robust classical information via "majority" is supported by the evolution.)
Showing (or assuming) that the
system is well approximated by a time-smoothed version may also be relevant.
For Navier--Stokes evolutions as for other classes 
of evolutions, proving or imposing ``no-computation''
%Deriving or imposing a bounded-degree restriction for a class of physical systems or considering a smoothed-stochastic evolutions for them
may also lead to additional interesting non-classical
conserved quantities.
%\subsubsection *{Free will}
%\subsubsection *{The debate about quantum computing}

%\section {Quid est noster computationis mundus?\protect\footnote{What is our computational world?}}
\section {What is our computational world?}

The remarkable progress witnessed during the past two decades in the field of experimental 
physics of controlled quantum systems places the decision between the pessimistic and optimistic 
hypotheses within reach. These two hypotheses 
reflect a vast difference in perspective regarding our computational world.   
Does the wealth of classical computations we witness in reality  
represent the full computational power that can be extracted from natural quantum physics processes, or
is it only the tip of the iceberg of a supreme computational power
used by nature and available to us?

Quantum computers represent
a new possibility acquired 
through a beautiful interaction of many scientific disciplines.
However unlikely it is and wherever it goes, this idea offers a terrific opportunity
and may change a great deal.
I expect that
the pessimistic hypothesis will prevail, yielding important outcomes
for physics, the theory of computing, and mathematics.
Our journey through probability distributions described by low-degree polynomials,
implicit modeling for noise, and error synchronization may provide some
of the pieces needed for solving the quantum computer puzzle.


\begin{thebibliography}{99}

\bibitem {Aar04} S. Aaronson,  Multilinear formulas and skepticism of quantum computing.
In {\it STOC '04: Proceedings of the thirty-sixth annual ACM symposium on
Theory of computing,} pp. 118--127. ACM Press, 2004.

\bibitem {AaAr13} S. Aaronson and A. Arkhipov,
The computational complexity of linear optics, {\it Theory of Computing} 4 (2013), 143--252.
arXiv:1011.3245


\bibitem {AhaBen97} D. Aharonov and M. Ben-Or,
Fault-tolerant quantum computation with constant error, in {\it STOC '97},
%(El Paso, TX),
ACM, New York, 1999, pp. 176--188.


%\bibitem {AHHH02} R. Alicki, M. Horodecki,
%P. Horodecki, and R. Horodecki, Dynamical description of
%quantum computing: generic nonlocality of quantum
%noise, {\it Phys. Rev. A} 65 (2002), 062101, quant-ph/0105115.

\bibitem {Ali04} R. Alicki, Quantum error correction fails
for Hamiltonian models, 2004, quant-ph/0411008.

%\bibitem  {ALZ06}
%R. Alicki, D.A. Lidar, and P. Zanardi, Are the assumptions of
%fault-tolerant quantum error correction internally consistent?,
%{\it Phys. Rev. A} 73 (2006), 052311, quant-ph/0506201.

\bibitem {AH06}
R. Alicki and M. Horodecki, Can one build a quantum hard drive? A no-go
theorem for storing quantum information in equilibrium systems, quant-ph/0603260.

\bibitem {Ali13}
R, Alicki, Critique of fault-tolerant quantum information processing,
in {\it Quantum Error Correction,} ed. D. A. Lidar and T. A. Brun, 
Cambridge University Press, 2013.



%\bibitem  {TB} B. B. Terhal and G. Burkard,
%Fault-tolerant quantum computation for local non-Markovian noise,
%{\it Phys. Rev. A} 71 (2005), 012336.



\bibitem {AMPS12} A. Almheiri, D. Marolf, J. Polchinski, and J. Sully, 
Black holes: complementarity or firewalls?, arXiv:1207.3123.




%${\rm http://www.}$${\rm wisdom.}$ ${\rm
%weizmann}.$${\rm ac.il/\string~ }$${\rm oded/on-qc.html}$,
%and S. Aaronson, Are quantum states exponentially long vectors?,
%2005, quant-ph/0507242.







\bibitem {AGP05} P. Aliferis, D. Gottesman, and J. Preskill,
Quantum accuracy threshold for concatenated
distance-3 codes, %2005, 
quant-ph/0504218.

\bibitem {AB09} S. Arora and B. Barak, {\it Computational Complexity:  Modern Approach}, Cambridge University Press, 2009.

\bibitem{Bar16}
A. Barvinok, Approximating permanents and hafnians of positive matrices, arXiv:1601.07518




\bibitem{BKS99}
I.~Benjamini, G.~Kalai, and O.~Schramm,
\newblock Noise sensitivity of Boolean functions and applications to
percolation,
\newblock {\em Publ. I.H.E.S.} 90 (1999), 5--43.

\bibitem {BGH13}
M. Ben-Or, D. Gottesman, A. Hassidim, Quantum refrigerator, arXiv:1301.1995.

\bibitem {BeVa93} E. Bernstein and U. Vazirani, Quantum complexity theory, 
{\it Siam J. Comp.} 26 (1997), 1411--1473. (Earlier version, {\it STOC}, 1993.)



\bibitem {BJS11}
 M. J. Bremner, R. Jozsa, D. J. Shepherd, Classical simulation of commuting quantum computations 
implies collapse of the
polynomial hierarchy, {\it Proc. Roy. Soc. A,}
467(2011)
%(2126)
,459--472, 2011.


\bibitem {BHL04} P. Busch, T. Heinonen, and P. Lahti, Noise and disturbance in quantum
measurement, {\it Phys. Lett. A} 320 (2004), 261--270.

\bibitem {DLS13} A. Daniely, N. Linial,  and S. Shalev-Shwartz, Complexity-theoretic
limitations on learning DNF's, arXiv:1311.2272.

\bibitem {Deu85}
D. Deutsch, Quantum theory, the Church-Turing principle and the
universal quantum computer, {\it Proc. Roy. Soc. Lond.} 
A 400 (1985), 96--117. 

\bibitem {Dya07}  M. I. Dyakonov, Is fault-tolerant quantum computation really possible? in: {\it 
Future Trends in Microelectronics. Up the Nano Creek, S. Luryi, J. 
Xu, and A. Zaslavsky (eds.)}, Wiley (2007), pp. 4--18.

\bibitem {Fey82} R. P. Feynman, Simulating physics with computers,
{\it Int. J. Theor. Phys.} 21 (1982), 467--488.

\bibitem {GaSt14} C. Garban and J. Steif, {\em Noise Sensitivity of Boolean Functions
and Percolation,} Cambridge University Press, 2014.

\bibitem {Gol04} O. Goldreich, On quantum computers, 2004, 2005, available 
on \href {http://www.wisdom.weizmann.ac.il/~oded/on-qc.html}{Goldreich's homepage}.

\bibitem {Gol08}
O. Goldreich, {\it Computetional Complexity:
A Conceptual Perspective,} Cambridge University Press, 2008.

\bibitem {Gol10} O. Goldreich, {\it P, NP, and NP-Completeness: The Basics of Complexity
Theory,} Cambridge University Press, 2010.

\bibitem {Got97} D. Gottesman, Stamilizer codes and quantum error-correction,
Ph. D. Thesis, Caltech, 1997.

\bibitem {Gur05} L. Gurvits, On the complexity of mixed 
discriminants and related problems, {\it Mathematical
Foundations of Computer Science 2005}, Lecture Notes in Computer Science, vol. 3618,
Springer, Berlin, 2005, pp. 447--458.

\bibitem {HaHa13} D. Harlow and P. Hayden, Quantum Computation vs. Firewalls, arXiv:1301.4504.

\bibitem {Har14} D. Harlow, Jerusalem lectures
on black holes and quantum information, arXiv:1409.1231.

\bibitem {HaRa96} S. Haroche and J. M.  Raimond, Quantum computing: dream or 
nightmare? {\it Phys. Today} 49 (1996), 51.

\bibitem {JSV04}
M. Jerrum, A. Sinclair and E. Vigoda, A polynomial-time approximation algorithm for
the permanent of a matrix with nonnegative entries, {\it Journal of the ACM} 51 (2004), 671--697

\bibitem {JLP14} S. P. Jordan, K. S. M. Lee, and J. Preskill, Quantum
algorithms for fermionic quantum
field theories
arXiv:1404.7115.


\bibitem {Kal05} G. Kalai, Thoughts on noise and quantum
computing, 2005, quant-ph/0508095.

\bibitem {Kal06} G. Kalai, How quantum computers can fail,
quant-ph/0607021.

\bibitem {Kal11} G. Kalai, How quantum computers fail:
quantum codes, correlations in physical systems, and noise
accumulation,
%\href{http://front.math.ucdavis.edu/1106.0485 }
arXiv:1106.0485.
%\href{http://simons.berkeley.edu/events/why-quantum-computers-cannot-work}{Vidotaped
%lectures, Simons Institute, 2013.}


\bibitem {Kal16} G. Kalai, The quantum computer puzzle, {\it Notices AMS} 63(2016), 508--516.


\bibitem{KalHar12} G. Kalai and A. Harrow, Debate over Lipton and Regan's blog {\it G\"odel's  Lost
 Letter and {\bf P = NP}},
\href{http://rjlipton.wordpress.com/2012/01/30/perpetual-motion-of-the-21st-century/}{post I},
\href{http://rjlipton.wordpress.com/2012/10/03/quantum-supremacy-or-classical-control/}{post VIII}.
and further discussion with Peter Shor, Harrow, and others over Kalai's blog {\it Combinatorics and more},
\href{https://gilkalai.wordpress.com/2013/03/13/a-few-slides-and-a-few-comments-from-my-mit-lecture-on-quantum-computers/}
{Post 1}, 
\href {https://gilkalai.wordpress.com/2014/03/18/why-quantum-computers-cannot-work-the-movie/}{Post 2}.

\bibitem{KalKin14} G. Kalai and G. Kindler, Gaussian noise sensitivity and BosonSampling, arXiv:1409.3093.

\bibitem {KalKup14} G. Kalai and G. Kuperberg, Contagious error sources would need time travel to prevent quantum computation, Physics Review A 92 (2015),
arXiv:1412.1907.

\bibitem {KeVa94} M. J. Kearns and U. V. Vazirani, {\it An Introduction to Computational 
Learning Theory}, M.I.T Press, 1994.


\bibitem{Kit97} A.~Y.~Kitaev, Quantum error
correction with imperfect gates, in {\it Quantum Communication,
Computing, and Measurement (Proc.\ 3rd Int.\ Conf.\ of Quantum
Communication and Measurement)}, Plenum Press, New York, 1997, pp. 181--188.


\bibitem {Kit03} A. Kitaev, Fault-tolerant quantum computation by anyons,
{\it Ann. Physics} 303 (2003), 2--30.

\bibitem{KLZ98} %E.~Knill, R.~Laflamme, and W.~H.~Zurek,
%Threshold accuracy for quantum computation,
%quant-ph/9610011;
%E.~Knill, R.~Laflamme, and W.~H.~Zurek,
%Resilient quantum computation, Science {\bf 279} (1998),
%342--345;
E.~Knill, R.~Laflamme, and W.~H.~Zurek, Resilient
quantum computation: Error models and thresholds, {\it Proc.\ Royal
Soc.\ London A }{454} (1998), 365--384, quant-ph/9702058.

\bibitem {lan95} R. Landauer, Is quantum mechanics useful?,
{\it Philos. Trans. Roy. Soc. London Ser. A} 353 (1995), 367--376.


\bibitem {Lev03}  L. Levin, The tale of one-way functions, {\it Problems of
Information Transmission (= Problemy Peredachi Informatsii)}
39 (2003), 92--103, cs.CR/0012023

\bibitem {MaSu13} J. Maldacena and L. Susskind, Cool horizons for entangled black holes,
arXiv:1306.0533.

\bibitem {Min78} H. Minc, {\em Permanents}, Addison-Wesley, 1978.

\bibitem {MR91} G. Moore and N. Read,
Nonabelions in the fractional quantum hall effect,
{\it Nuclear Physics B} 360 (1991), 362--393.

\bibitem {NC00} M. A. Nielsen and I. L. Chuang, {\it Quantum Computation
and Quantum Information}, Cambridge University Press, Cambridge, 2000.

\bibitem {Oza04} M. Ozawa, Uncertainty relations for joint measurements of noncommuting
observables, {\it Phys. Lett. A} 320 (2004), 367--374.

\bibitem {Pol14} L. Polterovich,  Symplectic geometry of quantum
noise, {\it Comm Math. Phys} 327 (2014), 481--519,
arXiv:1206.3707.

\bibitem {Pre98} J. Preskill, Quantum computing: Pro and con,
{\it Proc. Roy. Soc. London A} 454 (1998), 469--486, quant-ph/9705032.

\bibitem {Pre12} J. Preskill,
  Quantum computing and the entanglement frontier,  in {\it The Theory of the Quantum World: Proceedings of the
25th Solvay Conference on Physics,} ed. D. Gross, M. Henneaux,
and A. Severin, World Scientific, 2013, pp. 63--90,
arXiv:1203.5813.

\bibitem{Pre13} J. Preskill,
Sufficient condition on noise correlations for scalable quantum computing, 
{\it Quant. Inf. Comput.} 13 (2013), 181--194.


\bibitem{ShBe14} S. Shalev-Shwartz and S. Ben-David, {\it Understanding Machine Learning, From Theory to Algorithms,}
Cambridge University Press, 2014. 

\bibitem {Sho99} P.  W.  Shor, Polynomial-time
algorithms for prime factorization and
discrete logarithms on a quantum computer,
{\it SIAM Rev.} 41 (1999), 303--332. Eralier vesion, FOCS, 1994.


%(Earlier version,
%{\it Proceedings of the 35th Annual Symposium on Foundations of
%Computer Science}, 1994.)

\bibitem {Sho95} P. W. Shor, Scheme for reducing
decoherence in quantum computer
memory, {\it Phys. Rev. A} 52 (1995), 2493--2496.

\bibitem{Sho96} P. Shor, Fault-tolerant quantum computation,
in Proc. 37nd Annual Symposium on Foundations of Computer Science, 
IEEE Computer Society Press, pp. 56--65, 1996.

\bibitem {Sim94} D. R. Simon (1994), On the power of quantum computation, 
in {\it Proceedings of the 35th Annual Symposium
on Foundations of Computer Science,} IEEE Computer Society Press, Los Alamitos, CA, pp. 116--123.

\bibitem {Ste96}
A.~M.~Steane, Error-correcting codes in
quantum theory, {\it Phys.\ Rev.\ Lett.\ }{ 77} (1996), 793--797.


\bibitem  {TerBur05} B. B. Terhal and G. Burkard,
Fault-tolerant quantum computation for local non-Markovian noise,
{\it Phys. Rev. A} 71 (2005).

\bibitem {TeDi04}
B. Terhal and D. DiVincenzo, Adaptive quantum computation, constant depth quantum circuits 
and Arthur-Merlin games,
{\it Quant. Inf. Comp.} 4, 134--145 (2004).
arXiv:quantph/0205133.


\bibitem {TCF+12}
S. Trotzky, Y. Chen, A. Flesch, I. P. McCulloch, U. Schollw\"ock, J. Eisert, and I. Bloch,
Probing the relaxation towards equilibrium in an isolated strongly correlated 1D Bose gas, 
{\it Nature Physics} 8, 325 (2012), arXiv:1101.2659.

\bibitem {Tao15} T. Tao, Finite time blowup for an averaged three-dimensional Navier--Stokes equation,
 Jour. AMS (2015),   arXiv:1402.0290. 

\bibitem {TrTi96} L. Troyansky and N. Tishby, Permanent uncertainty: On the quantum evaluation
of the determinant and the permanent of a matrix, in {\it Proc. 4th Workshop on Physics and Computation}, 1996.

\bibitem {Unr95} W. G. Unruh, Maintaining coherence in quantum computers,
{\it Phys. Rev. A} 51 (1995), 992--997.

\bibitem{War13} R. H. Warren, Numeric experiments on the 
commercial quantum computer, {\it Notices AMS} 60 (2013) 1434--1438.

\bibitem {Yao93} A. Yao (1993), Quantum circuit complexity, in {\it Proceedings of the 34th 
Annual Symposium on Foundations of Computer Science,} IEEE Computer Society Press, Los Alamitos, CA, pp. 352--361.

\bibitem {Yao03} A.C.-C. Yao,  Classical physics and the Church-Turing Thesis, {\it Jour. ACM}  50 (2003), 100--105.

\end{thebibliography}
\end {document}